\documentclass[12pt]{article}
\usepackage{latexsym,graphicx}
\newlength{\zeichenbreit}
\newlength{\slashbreit}
\newcommand{\be}{\begin{equation}}
\newcommand{\ee}{\end{equation}}
\newcommand{\bea}{\begin{eqnarray}}
\newcommand{\eea}{\end{eqnarray}}
\newcommand{\pkt}{\; .}
\newcommand{\kma}{\; ,}
\newcommand{\call}{{\cal L}}
\newcommand{\calm}{{\cal M}}
\newcommand{\calf}{{\cal F}}
\newcommand{\cale}{{\cal E}}
\newcommand{\calc}{{\cal C}}
\newcommand{\calv}{{\cal V}}
\newcommand{\bfx}{{\bf x}}

\newcommand{\bfk}{{\bf k}}
\newcommand{\eqn}[1]{(\ref{#1})}
\newcommand{\intk}[1]{\int\!\frac{d^3 k}{(2\pi)^3#1}\,}

\newcommand{\re}{{\rm Re}}

\begin{document}

\begin{titlepage}
\begin{flushright}
DO-TH-07/02 \\
hep-th/0702009 \\
January 2007
\end{flushright}

\vspace{20mm}
\begin{center}
{\Large \bf
Quantum back-reaction of the superpartners in a large-$N$ 
supersymmetric hybrid model}
\vspace{10mm}

{\large
J\"urgen Baacke \footnote{e-mail: baacke@physik.uni-dortmund.de},
Nina Kevlishvili  
\footnote{e-mail: nina.kevlishvili@het.physik.uni-dortmund.de},
and Jens Pruschke
\footnote{e-mail: jens.pruschke@uni-dortmund.de}}

\vspace{15mm}

{\large  Institut f\"ur Physik, Universit\"at Dortmund \\
D - 44221 Dortmund, Germany
}

\vspace{15mm}

\bf{Abstract}
\end{center}
We study the supersymmetric hybrid model near and after the end
of inflation. As usual, we reduce the model to a purely scalar
hybrid model on the level of the classical fields.
But on the level of quantum fluctuations and their backreaction
we take into account  all superpartners of the waterfall field 
in a large-$N$ approximation. The evolution after slow roll displays
two phases with a different characteristic behaviour of the
classical and fluctuation fields.  We find that
the fluctuations of the pseudoscalar superpartner are of particular
importance in the late time phase. The motion of the waterfall field towards its
classical expectation value is found to be very slow and suggests
a rather flat potential and a stochastic force. 
\end{titlepage}

\setcounter{page}{2}

\section{Introduction}\label{intro}
The hybrid model has been proposed by Linde  
\cite{Linde:1990gz,Linde:1991km,Linde:1993cn} as a possible inflationary
scenario, several variants of the model have been discussed 
recently 
\cite{Copeland:1994vg,Garcia-Bellido:1997wm,Micha:1999wv,
Buchmuller:2000zm,Nilles:2001fg,Asaka:2001ez,Cormier:2001iw}.
Here we consider the model in the context of preheating after inflation,
and not inflation itself. This period has many interesting aspects
and has received a wide attention
\cite{Garcia-Bellido:1997wm,Micha:1999wv,
Buchmuller:2000zm,Nilles:2001fg,Asaka:2001ez,
Cormier:2001iw,Lyth:1998xn,Garcia-Bellido:1999sv,
Bastero-Gil:1999fz,Krauss:1999ng,Felder:2000hj,Felder:2001kt,
Copeland:2001qw,Garcia-Bellido:2002aj,Borsanyi:2002tm,Borsanyi:2003ib,
Alabidi:2006hg,Barnaby:2006km}.
The hybrid  model may arise naturally in the context of
supersymmetry and supergravity \cite{Dvali:1994ms,Copeland:1994vg,
Lyth:1998xn,Bastero-Gil:1999fz,
Micha:1999wv,Buchmuller:2000zm,Asaka:2001ez}.
Another aspect of the type of model investigated here may be that
it simulates a second order phase transition, where the change in the
effective potential does not arise from a decrease in temperature but
is mediated by an effective field. It thereby replaces models
\cite{Boyanovsky:1996rw,Bowick:1998kd,Felder:2000hj,Felder:2001kt} 
where a rapid decrease of  the temperature is simulated by an 
instantaneous quench.

The supersymmetric hybrid model \cite{Dvali:1994ms,Copeland:1994vg}
is often reduced to a purely scalar
hybrid model. This is generally done on the level of classical
fields. Loop effects are invoked sometimes in order to generate an 
effective potential for the inflaton field, replacing 
an explicit mass term. In the period right after inflation,
at the time when the phase transition of the waterfall field takes place,
and the subsequent preheating period, quantum fluctuations are produced
abundantly, by spinodal decomposition or by parametric
resonance, and their backreaction has to be taken into account.
In previous publications  
\cite{Bastero-Gil:1999fz,Cormier:2001iw,Baacke:2003bt}
the quantum fluctuations of both the inflaton and the waterfall field
and their backreaction on the classical fields were discussed
in different approximations. 
Here we would like to investigate the r\^ole of the fluctuations of the
superpartners, as appropriate for a supersymmetric model. The first 
studies of a supersymmetric quantum field theory out-of-equilibrium 
have appeared some years ago \cite{Baacke:2001eg,Baacke:2000jp}, but 
they have not been put into the context of cosmology. 
Here we consider the supersymmetric hybrid model which is one of
the standard multifield models of inflation.

In a model with a spinodal instability
it is essential to include the backreaction of the modes onto themselves.
It is this backreaction which stabilizes the system by
shifting the negative squared masses back to positive values. Such a 
backreaction can be included in the most simple form either in the
large-$N$ or the Hartree approximations. The Hartree approximation would be 
much more involved; also, it is not clear how to treat the fermion fields
consistently.  Here we use, as a first approach to the problem, 
a large-$N$ approximation, along the lines of 
\cite{Baacke:2001eg,Baacke:2000jp}. The original model contains an
$U(1)\simeq SO(2)$ symmetry for the waterfall field, which we elevate to
an $SO(N)$ symmetry as a basis to the large-$N$ approximation.  

The plan of the paper is as follows. 
In section \ref{basics} we present the basic model, its large-$N$
version and the dynamical equations in unrenormalized form. 
Renormalization is discussed in Appendix \ref{renormalization}.
The numerical approach and the choice of parameters are discussed
in section \ref{numsandparams}. The results of the
numerical simulations are presented and discussed in section
\ref{numericalresults}. We end with a summary and conclusions in
section \ref{summaryandconclusions}.


\section{The model and basic dynamical equations}
\label{basics}
The  supersymmetric hybrid model  
\cite{Dvali:1994ms,Garcia-Bellido:2002aj}
is usually based on  the superpotential
\be \label{superpotential}
W(S,\Phi_1,\Phi_2)=\kappa S(\Phi_1\Phi_2-\mu^2/\kappa)
\pkt \ee
By various arguments the model is then reduced to the standard
hybrid model involving two fields:  the inflaton field $\phi$,
as the scalar field of the supermultiplet $S$, and the ``waterfall'' field 
$\chi$ as the remnant of the two superfields $\Phi_1$ and $\Phi_2$.
The Lagrangean takes the form of the standard hybrid model
\be
\call=\frac{1}{2}\partial_\mu\phi\partial^\mu\phi
+\frac{1}{2}\partial_\mu\chi\partial^\mu\chi
-\frac{1}{2}m^2\phi^2-\frac{1}{2}g^2\phi^2\chi^2\nonumber 
-\frac{\lambda}{4}(\chi^2-v^2)^2
\pkt\ee
The only aspect of this Lagrangean that links it to the supersymmetric
model  is the relation between the couplings $\lambda$ and $g$:
\be
 \lambda  =\frac{g^2}{2}=\kappa^2
\pkt\ee
The vacuum expectation value $v$ is given by $v=\mu/\sqrt{\kappa}$. 
Furthermore the mass $m$ is often taken to be zero and the potential for
the field $\phi$ is generated by the quantum fluctuations of the waterfall
supermultiplet.
In Ref. \cite{Baacke:2003bt} 
A. Heinen and one of us (J.B.) have considered this purely 
scalar hybrid model
including the one-loop quantum backreaction in the Hartree
approximation. However: even if we reduce the supersymmetric model
to the scalar hybrid model on the level of classical fields, there are more
quantum fluctuations than just those of the two scalar fields, namely
the quantum fluctuations of the pseudoscalar and fermionic superpartners.
These are dismissed in the usual treatment of the model, and it their 
impact that we want to study.

When taking into account the quantum fluctuations in a theory with
spontaneous symmetry breaking one has to go beyond
the one-loop approximation: the squared masses of the fluctuations 
can become negative, and this leads
to a fast breakdown of the system due to an exponential increase of the
fluctuations. In the large-$N$ and Hartree aproximations the backreaction
of these modes onto themselves stabilizes the evolution, and this seems
to be a sound feature of these approximations. The Hartree approximation, when
applied to all fields becomes quite involved, and there are conceptual
problems with incorporating the fermion fields.

An approximation that has already been formulated for a supersymmetric
system is the large-$N$ approximation \cite{Baacke:2000jp,Baacke:2001eg}. 
Here we will consider 
this approximation, but the original model with the superpotential
\eqn{superpotential} does not display any large $N$ in which we
could expand. But of course there may be generalizations like the
one discussed in Ref. \cite{Dvali:1997wz} where the inflaton couples
to a field with a higher symmetry group.
Specifically, we will consider here a large-$N$ extension of the 
original model, based on the potential
\be
W(\Phi_0,\Phi_1,\dots,\Phi_N)=
\frac{\kappa}{\sqrt{N}} 
\Phi_0\left(\sum_{k=1}^N\Phi_k^2-\frac{N}{2} v^2\right)
\pkt\ee
The previous model is recovered for $N=2$ by the identification
\bea
\Phi_1 &\to& \frac{1}{\sqrt{2}}(\Phi_1+i\Phi_2)=\Phi \kma
\\
\Phi_2 &\to& \frac{1}{\sqrt{2}}(\Phi_1-i\Phi_2)=\bar \Phi
\pkt\eea
The latter notation is used in the work of Dvali et al. \cite{Dvali:1994ms}.
Note that all fields are complex superfields. The original model has a
$U(1)\simeq SO(2)$ symmetry $\Phi \to \exp(i\gamma)\Phi$, 
$\bar \Phi \to \exp(-i\gamma)\bar\Phi$ which here is converted 
into an $SO(N)$ symmetry.

The auxiliary fields are given by
\bea
F_0^*(\varphi)&=&-\frac{\partial W}{\partial \Phi_0}
(\varphi)=-\frac{\kappa}{\sqrt{N}}\left(
\sum_{k=1}^N\varphi_k^2-\frac{N}{2}v^2\right) \kma
\\
F_k^*(\varphi)&=&-\frac{\partial W}{\partial \Phi_k}(\varphi)=-2\frac{\kappa}{\sqrt{N}} 
\varphi_0\varphi_k
\pkt\eea
 The scalar potential becomes
\be\label{pot}
V=\sum_{i=0}^N|F_i|^2=\frac{\kappa^2}{N}\left|\sum_{k=1}^N
\varphi_k^2-\frac{N}{2}v^2\right|^2+4\frac{\kappa^2}{N}|\varphi_0|^2
\sum_{k=1}^N|\varphi_k|^2
\pkt\ee
The Yukawa part of the fermion Lagrangian is given by
\bea \nonumber
\call_{\rm Y}&=&-\frac{\kappa}{2\sqrt{N}}\sum_{k=1}^N\bar\psi_0
\left[\varphi_k(1-\gamma_5)+\varphi_k^*(1+\gamma_5)\right]\psi_k
\\&&-\frac{\kappa}{2\sqrt{N}}\sum_{k=1}^N\bar\psi_k
\left[\varphi_k(1-\gamma_5)+\varphi_k^*(1+\gamma_5)\right]\psi_0
\\\nonumber
&&-\frac{\kappa}{2\sqrt{N}}\sum_{k=1}^N\bar\psi_k
\left[\varphi_0(1-\gamma_5)+\varphi_0^*(1+\gamma_5)\right]\psi_k
\pkt\eea

The large-$N$ limit is obtained by introducing
two spatially homogeneous classical fields $\phi_0$ 
and $\phi_1$, and $N+1$ fluctuation
fields $\eta_i$ via
\bea\nonumber
\varphi_0&=&\sqrt{\frac{N}{2}}\phi_0(t)+\eta_0 \kma
\\
\varphi_1&=&\sqrt{\frac{N}{2}}\phi_1(t)+\eta_1 \kma
\\\nonumber
\varphi_k&=&\eta_k \;\;\; k=2..N
\pkt \eea
The classical fields are taken to be real.

The large-$N$ part of the bosonic Lagrangian is given by
\bea\nonumber
&&\frac{1}{N}\call_{\rm{N,bos}}=\frac{1}{2}\dot \phi_0+
\frac{1}{2}\dot\phi_1^2
+\frac{1}{N}\sum_{k=2}^N\partial_\mu\eta^*\partial^\mu\eta-\kappa^2\phi_0^2\phi_1^2
-2\kappa^2\phi_0^2\frac{1}{N}\sum_{k=2}^N|\eta_k|^2
\\&&-\frac{\kappa^2}{4}\left(\phi_1^2-v^2\right)^2
-\kappa^2\left(\phi_1^2-v^2\right)\frac{1}{N}
 \sum_{k=2}^N\re\eta_k^2
-\frac{\kappa^2}{N^2}\left|\sum_{k=2}^N\eta_k^2\right|^2
\kma \eea
while the fermionic part becomes
\be
\frac{1}{N}\call_{\rm N,ferm}=
\frac{1}{N}\sum_{k=2}^{N}\frac{1}{2}\bar\psi_k\left(i\gamma_\mu\partial^\mu-
\kappa\sqrt{2}\phi_0\right)\psi_k \pkt
\ee 
Note that in leading order of large-$N$ we only consider the 
$N-1$ ``transversal''
quantum fluctuations $\eta_k,\psi_k, k=2..N$. 

We now introduce the real fields $a_k,b_k$.
\be
\eta_k =\frac{1}{\sqrt{2}}(a_k+ib_k)\;\;\; k=2..N
\pkt \ee

The part quadratic in the bosonic quantum fluctuations becomes
\bea \nonumber
\call_{\rm N, fluct}&=&
\frac{1}{2}\sum_{k=2}^N \left[\partial_\mu a_k\partial^\mu a_k+
\partial_\mu b_k \partial^\mu b_k-\kappa^2\phi_0^2 (a_k^2+b_k^2)
\right.
\\
&&\left. -\frac{\kappa^2}{2}(\phi_1^2-v^2)(a_k^2-b_k^2)\right]
\pkt\eea
 So on the tree level  the masses are given by
\bea
m_a^2&=&2\kappa^2\phi_0^2+\kappa^2(\phi_1^2-v^2) \kma
\\
m_b^2&=&2\kappa^2\phi_0^2-\kappa^2(\phi_1^2-v^2) \kma
\\
m_f^2&=&2\kappa^2\phi_0^2
\eea
satisfying  the supersymmetry relation
\be
\sum (-1)^{N^f_i}m_i^2=0
\pkt\ee

We now introduce a Gaussian wave functional, so that
\bea
<a_i(x)a_j(y)>&=&<a(x)a(y)>\delta_{ij} \kma
\\
<b_i(x)b_j(y)>&=&<b(x)b(y)>\delta_{ij} \kma
\\
<a_i(x)b_j(y)>&=&0 \kma
\\
<\bar\psi_i(x)\psi_j(y)>&=&<\bar\psi(x)\psi(y)>\delta_{ij}
\eea
and all the higher correlation functions reduce to the two-point ones.
We now can replace 
\be
\sum_k<a_ka_k>=(N-1)<a^2>\simeq N <a^2>
\ee
and similarly for all terms which are second order in the
fluctuations. The quartic term needs some care:
we have
\be
<\sum_{k,l} (a_k^2+2ia_kb_k-b_k^2)(a_l^2+2ia_lb_l-b_l^2)>
=N^2(<a^2>-<b^2>)^2+ O(N)
\ee 
as for correlations between some $a_k$ and some $a_l$ we get a
Kronecker $\delta_{kl}$ and for such correlations only one summation remains,
yielding a factor $N$ instead of $N^2$.

Using the Gaussian factorization and treating all transverse fluctuations as
dynamically identical we obtain the bosonic part of the Lagrangian:
\bea\nonumber
\frac{1}{N}\call_{\rm N, bos}&=&
\frac{1}{2}\dot \phi_0^2+
\frac{1}{2}\dot\phi_1^2 -\kappa^2\phi_0^2\phi_1^2-
\frac{\kappa^2}{4}\left(\phi_1^2-v^2\right)^2
\\\nonumber
&&+\frac{1}{2}(\partial_\mu a\partial^\mu a+\partial_\mu b
\partial^\mu b)
-\kappa^2\phi_0^2 (a^2+b^2)
-\frac{\kappa^2}{2}(\phi_1^2-v^2)(a^2-b^2)
\\
&&-\frac{\kappa^2}{4}(a^2-b^2)^2
\kma \eea
while the fermion part takes the form
\be
\frac{1}{N}\call_{\rm N, ferm}=
\frac{1}{2}\bar\psi\left( i\gamma_\mu\partial^\mu - \kappa\sqrt{2}\phi_0
\right)\psi
\pkt\ee

The effective masses now become
\bea \label{amass}
\calm_a^2&=&2\kappa^2\phi_0^2+\kappa^2(\phi_1^2-v^2)+\kappa^2\left(<a^2>-<b^2>\right)\kma
\\\label{bmass}
\calm_b^2&=&2\kappa^2\phi_0^2-\kappa^2(\phi_1^2-v^2)-\kappa^2\left(<a^2>-<b^2>\right)\kma
\\\label{fmass}
\calm_f^2&=&2\kappa^2\phi_0^2\kma
\eea
or
\bea
\calm_a^2&=&\calm_f^2+\calm_-^2 \kma
\\
\calm_b^2&=&\calm_f^2-\calm_-^2
\pkt\eea
The supersymmetry sum rule for the masses is still satisfied.

The mass $\calm_-^2$ satisfies a gap equation
\be
\calm_-^2=\kappa^2(\phi_1^2-v^2)+\kappa^2\left(<a^2>-<b^2>\right)
\kma \ee
where the right hand side is a functional of $\phi_0,\phi_1$ and
$\calm_-^2$. The gap equation has to be solved at $t=0$, later
on it is then satisfied automatically. We define 
\be
m_{j0}^2=\calm_j^2(0)\;\;\; j=a,b,f
\ee
and the potentials
\be
\calv_j(t)=\calm_j(t)-m_{j0}^2 
\pkt
\ee
The equations of motion for the classical fields are obtained as
\bea
\ddot \phi_0+2\kappa^2\phi_0\phi_1^2+2\kappa^2\phi_0(<a^2>+<b^2>)+
\frac{1}{\sqrt{2}}\kappa<\bar\psi\psi>&=&0 \kma\\
\ddot\phi_1+2\kappa^2\phi_0^2\phi_1+\kappa^2\phi_1(\phi_1^2-v^2)
+\kappa^2\phi_1(<a^2>-<b^2>)&=&0
\pkt\eea

The second equation may also be written as
\be
\ddot \phi_1 +\calm_a^2\phi_1 =0
\pkt\ee
Of course it remains a nonlinear equation and the equations for
$\phi_0$ and $\phi_1$ form a coupled system.

The fluctuations satisfy
\bea
\ddot a -\Delta a +\calm_a^2 a&=&0\kma
\\
\ddot b - \Delta b +\calm_b^2 b &=&0\kma
\\
(i\gamma^\mu\partial_\mu-\sqrt{2}\kappa\phi_0)\psi&=&0
\pkt\eea
Unlike the case of the hybrid model in the Hartree 
approximation the system of equations for the different 
fluctuations are uncoupled in the large-$N$
limit. This would change if $\phi_1$ were allowed to be complex.

We expand the fluctuations into mode functions as
\bea
a(\bfx,t)&=&\int \frac{d^3 k}{(2\pi)^3}
(a(\bfk)f^a_k(t) e^{i\bfk\cdot \bfx}+a^\dagger(\bfk)f^{a*}_k(t) 
e^{-i\bfk\cdot \bfx})\kma
\\
b(\bfx,t)&=&\int \frac{d^3 k}{(2\pi)^3}
(b(\bfk)f^b_k(t) e^{i\bfk\cdot \bfx}+b^\dagger(\bfk)f^{b*}_k(t)
 e^{-i\bfk\cdot \bfx})\kma
\\
\psi(\bfx,t)&=&\sum_s\int \frac{d^3k}{(2\pi)^3}\left[
c(\bfk,s)U_s(\bfk,t)+c_\dagger(-\bfk,s) V_s(\bfk,s,t)\right]
e^{i\bfk\cdot\bfx}
\pkt\eea
The mode functions $f^{a/b}_k(t)$ satisfy
\be
\ddot f^{a/b}_k +k^2f^{a/b}_k  +\calm_{a/b}^2 f^{a/b}_k=0
\ee
and have the initial conditions
\bea
 f^{a/b}_k(0)&=&1\kma
\\
 \dot f^{a/b}_k(0)&=&-i\omega_{a/b0}\kma
\eea
with $\omega_{j0}= \sqrt{k^2+m^2_{j0}}, j=a,b,f$.

We write the spinors $U_s(\bfk,t)$ as
\bea
U_s(\bfk,t)=N_0\left[i\partial_t+ 
\tilde{\cal H}_\bfk(t)\right]f^\psi_k(t)
\left(\begin{array}{c}
\chi_s \\ 0
\end{array}\right)\kma
\\
V_s(\bfk,t)=N_0\left[i\partial_t+ 
\tilde{\cal H}_{-\bfk}(t)\right]g^\psi_k(t)
\left(\begin{array}{c}
0 \\ \chi_s
\end{array}\right)\kma
\eea
with the Fourier-transformed Hamiltonian
\be
\tilde{\cal H}_\bfk(t)=\mbox{\boldmath $\alpha k$\unboldmath} +\calm_f(t)\beta
\pkt \ee
For the two-spinors $\chi_s$ we use helicity eigenstates, i.e.,
\be
\mbox{\boldmath$ \hat k\sigma$\unboldmath} \chi_\pm=\pm \chi_\pm
\pkt
\ee
The mode functions $f^\psi_k(t)$ and $g^\psi_k(t)$ 
only depend on $k=|\bfk|$;
they obey the second order differential equations
\bea\label{fsec}
\left[
\frac{d^2}{dt^2}-i\dot\calm_f(t)+ k^2+\calm_f^2(t)
\right]f^\psi(k,t)&=&0\kma \\\label{gsec}
\left[
\frac{d^2}{dt^2}+i\dot \calm_f(t)+k^2+\calm_f^2(t)
\right]g^\psi_k(t)&=&0\pkt
\eea
The initial conditions are
\bea\label{fginit}
f^\psi_k(0)=1&,& \dot f^\psi_k(0)=-i\omega_{f0}\kma
\\
g^\psi_k(0)=1&,& \dot g^\psi_k(0)=i\omega_{f0}\kma
\eea
so that $g^\psi_k(t)=f^{\psi*}_k(t)$.

Then the mode sums or fluctuation integrals take the 
form
\bea
<a^2>&=&\intk{2\omega_{a0}}|f^a_k(t)|^2 \kma
\\
<b^2>&=&\intk{2\omega_{b0}}|f^b_k(t)|^2 \kma
\\
<\bar\psi\psi> &=&-2\intk{2\omega_{f0}}
\left\{
2\omega_{f0}-\frac{2\bfk^2}{\omega_{f0}+m_{f0} }|f^\psi_k(t)|^2
\right\}\pkt
\eea 

All the fluctuation integrals are divergent. The procedure to separate
them into finite integrals over the mode functions and renormalized 
finite parts of the divergent leading order contributions
has been described in Refs.\cite{Baacke:1996se,Baacke:1998di}.
 The renormalization is discussed in 
Appendix \ref{renormalization}.


\section{Numerical  procedure and choice of parameters}
\label{numsandparams}
We have implemented the equations of motion derived in the
last section into a numerical code. After separation of the
divergent parts and renormalization the computation of
the fluctuation integrals reduces to finite integrals and
is straightforward. The procedure is described in detail in
Refs. \cite{Baacke:1996se,Baacke:1998di}. The choice of the momentum grid
and of the time steps depends of course on the parameters of the model
and on the initial conditions. For our choices (see below) we have
typically extended the momentum integration up to $p_{\rm max}=15$ and used
a grid of $700$ momenta, slightly concentrated towards $p=0$. 
The reliability was checked using energy conservation
and by the constancy of the Wronskians.  With small values of $\kappa$ 
the time evolution is very slow. Nevertheless the integration of the
mode functions at high momenta requires sufficiently small time steps.
We have chosen $\Delta t \simeq 10^{-4}$ and followed the evolution up
to times of $t_{\rm max}\simeq 60000$. 

The original model has two parameters, $\kappa$ and $\mu$. We have chosen
$\mu=1$, which then defines the mass and inverse length scale.
Then $v=\mu/\sqrt{\kappa}=1/\sqrt\kappa$. This fixes the units we 
have chosen for the plots. The renormalization scale 
was taken to be $m=\kappa v=\mu\sqrt{\kappa}$.
For each run we have to specify two more parameters, $\phi_0(0)$ and
$\phi_1(0)$. We have to give the waterfall
field a small initial value because otherwise it remains zero
forever. The value cannot be chosen ``infinitesimally small'' like
$10^{-7}$, as in the absence
of Hubble expansion and of an explicit inflaton mass, the system remains at
$\phi_0=\phi_0(0)$ for a very long time. The fluctuation integrals
are nonzero at $t=0$ and can in principle initiate a
time evolution of $\phi_0$, but for small initial
excitations their numerical values are too small and for larger ones 
there is a cancellation between bosonic and fermionic contributions
to the equation of motion of $\phi_0$. Typically we have chosen
$|\phi_1(0)|\simeq 0.01$.

As to the parameter $\kappa$ a value of $\kappa=0.01$ has been
suggested in Ref. \cite{Dvali:1994ms}, much smaller values have been
discussed in Ref. \cite{Senoguz:2004vu}, the simulations in Ref. 
\cite{Bastero-Gil:1999fz} were presented for $\kappa=0.001$.
 In any case $\kappa$ is a small parameter, and we have performed simulations
with $\kappa=0.1,~ 0.01$ and $0.001$. As the time evolution gets slower
with decreasing values of $\kappa$ we present most of our results
for $\kappa =0.1$. The qualitative features remain the same
for smaller values, but a detailed study would require very long 
CPU times.


\section{Numerical results}
\label{numericalresults}
\subsection{Time evolution I: the slow roll period.}

This period takes a time $t_1$ which ranges between $50$ for
$\kappa=0.1$ and low excitations and several thousand for $\kappa=0.001$
and large initial value of $\phi_0$. The evolution is mostly classical,
the fluctuations remain very small. 
This period ends once the mass squared of the scalar ($a$ field) fluctuations 
gets negative and the system enters the spinodal
regime; this is the usual end of the slow roll phase, marked
in the figures as a vertical line at $t=402$  for $\phi_0(0)=4$ and
at $t=1374$ for $\phi_0(0)=64$.
The behaviour of the masses in this transition region is displayed
in Figs. \ref{figure:masszbs} and \ref{figure:masszbl} for
small and large initial exctitations, respectively. 
$\calm_a^2$ gets negative first, 
then the strong increase in the $a$ field fluctuations 
drives $\calm_b^2$ to negative values. Subsequently
$\calm_a^2$ and $\calm_b^2$ oscillate around zero
with alternating signs. After a few oscillations $\calm_b^2$ 
remains positive while 
$\calm_a^2$ continues between positive and negative values.
We have entered the second stage of evolution.

The point where $\calm_a^2$ becomes negative marks the
end of inflation. At this point the energy density 
$\cale$ of the classical
fields is transferred to the quantum fluctuations, and the pressure,
which  was essentially equal to $-\cale$ in the inflationary phase
becomes the one of a massive or massless gas (see e.g. 
\cite{Boyanovsky:1996sq} of an out-of-equilibrium
analysis). This transfer of energy density can be seen in
Figs. \ref{figure:energysm} and \ref{figure:energylm} for small and large
initial excitiations, respectively, while it takes a very long time
for small excitations. We will see, however, that  
the evolution of the classical fields towards the classical minimum 
of the potential will take a long time and is not as instantaneous
as often assumed.

If the Hubble expansion were taken into account, the evolution
during the slow roll stage would be modified essentially. So our 
numerical results for this period are not really relevant, they 
mostly set an initial condition
for the second stage, which is characterized by the emergence
of the quantum fluctuations. 
 

\subsection{Time evolution II: the intermediate period.}

In the intermediate period there are two qualitatively different
kinds of evolution. For small excitations $\phi_0(0)$ the 
waterfall field $\phi_1$ decides to move either into the positive 
or the negative direction and attains
some average value $\bar\phi_1$ whose absolute value is smaller than
the tree level vacuum expectation value $v$. We call this the 
``broken symmetry phase''.
For large excitations $\phi_0$ the waterfall field oscillates around
zero. We call this the ``symmetric phase''. Such a phenomenon has
similarly observed in Ref. \cite{Baacke:2003bt}, we discuss it
in more detail below.
We display the behaviour of the classical fields
during this second and the early third period 
in Figs. \ref{figure:phism2} and
\ref{figure:philm2} for low and high initial excitation, respectively.

The quantum fluctuations start developing
right after entering the spinodal regime. 
The fermion fluctuations remain very small throughout. 
The $a$ amplitudes have initially developed exponentially in a
low momentum band and $\calf_a$ has reached large values right away.
Subsequently these fluctuations develop only very slowly, though 
$\calm_a^2$ oscillates around zero and therefore becomes negative
periodically. This entails an exponential
evolution in the low momentum band.
However, the periods of exponential evolution
alternate with periods where $\calm_a^2>0$; these intermittent
periods modify the initial conditions for the next exponential evolution.
So we have neither parametric resonance nor a spinodal evolution and the 
fluctuation integral $\calf_a$ does not evolve significantly.

The squared mass of the $b$ field 
fluctuations oscillates  around some positive 
value. So they could develop by parametric resonance. A resonance peak
can indeed be seen at the beginning of, and during the second
period. However, the amplitude of the mass oscillations is  small
and is of course not cleanly periodic. So the resonance does not evolve
efficiently.

We display the evolution of the fluctuation integrals during the second and
early third period in Figs. \ref{figure:fluctsm2} and \ref{figure:fluctlm2},
for small and large initial values of $\phi_0$, respectively.

Right after the intermediate period the structure of the
momentum spectra still displays the same features as in the early
stage of this period. They just have become somewhat irregular. They are
displayed in Figs. \ref{figure:spectrsabm} and \ref{figure:spectrlabm}
for small and large initial values of $\phi_0$.


\subsection{Time evolution III: The late time behaviour}

The intermediate period ends with a second strong evolution
of the quantum fluctuations, this time including those of the
$b$ fields. With the decrease of the field $\phi_0$, 
the increase of $\phi_1$ and of $<a^2>$
the squared mass $\calm_b^2$ is, at some point, driven again to 
negative values,
see Eq. \eqn{bmass}. The time at which this occurs is marked by a vertical
line, at $t=1733$ for $\phi_0(0)=4$ and $t=2438$ for $\phi_0(0)=64$.
Of course the increase of the $b$ fluctuations drives 
the $b$ mass back to positive values, but then the $a$ mass gets
negative again, and so on. This behaviour at the onset of the late time
regime is displayed in Figs. \ref{figure:masszel}. The vertical line marks
the time where $\calm_b^2$ starts taking negative values again.

Now the $b$ field 
fluctuations develop significantly. This entails
strong fluctuations in all squared masses and a characteristic
change in the evolution of the waterfall field. While in the 
intermediate period it keeps oscillating more or less regularly around
zero (symmetric phase, high excitations) or some finite value
(broken symmetry phase, low excitations), it now starts shifting towards
the classical minimum $\phi_1=v$ in quite an irregular motion. The shift 
is very slow and the motion does not display any periodic oscillations. 
So the out-of-equilibrium effective potential 
(if at all one  may use such a term) seems to be quite flat, 
and the motion seems
to be determined by stochastic forces instead of  well-defined harmonic forces.
This is displayed in Figs. \ref{figure:phisl} and \ref{figure:phill}.
The stochastic behaviour may originate on the one hand from a
parametric resonance that is strongly disturbed by the presence and variation
of different time scales. On the other hand the oscillations 
of $\calm_a^2$ and $\calm_b^2$ around zero naturally lead to a 
diffusion process by the alternation of oscillating and exponential
time evolution.

In this late time period both $\calm_a^2$ and $\calm_b^2$ oscillate around
zero. So, for the low momentum modes the effective frequency
takes real and imaginary values in intermittent
time intervals. As before for the fluctuations of the
$a$ field we get now for both scalar fields a process 
where an exponential behaviour alternates with 
an oscillating evolution, which changes initial conditions for the
next exponential evolution. This time the amplitudes of oscillation
for $\calm_{a/b}^2$ are larger, the process becomes more effective 
and leads to a strong increase of the fluctuation integrals, as mentioned
above. At the same time the momentum spectra loose entirely their 
characteristic features related to
parametric resonance and/or spinodal evolution. They are concentrated 
at small momenta and fall off rapidly with momentum. These spectra
are displayed in  Fig. \ref{figure:spectrsabl} for $\phi_0(0)=4$ 
and in Fig. \ref{figure:spectrlabl} for $\phi_0(0)=64$.

This third period of evolution shows novel features that deserve to
be analyzed in more detail. Without the pseudoscalar superpartner,
the system remains in the  ``symmetric'' and ``broken symmetry'' phases
\cite{Baacke:2003bt}, here the waterfall field always evolves 
towards of one of its classical minima at late times. The stochastic
behaviour contrasts with the spinodal or parametric resonance
regimes and seems to represent a new kind of behaviour specific
of multifield quantum systems.


\subsection{ A phase transition}
\label{phasetransition}
As we have discussed previously the system behaves in two different
ways in the intermediate stage:
for small excitations the field $\phi_1$ shifts towards the direction of 
one of the classical minima and oscillates in an irregular way around some
mean value   which depends on the initial excitation. With increasing 
initial excitation this mean value becomes smaller 
and above some critical value
it becomes zero. This can be seen as a phase transition, with a symmetric
phase for large excitation, i.e. large energy densities.
We plot, in Fig. \ref{figure:phasetrans}
 
this average value as a function of $\phi_0(0)$, for $\kappa=0.1$.
 As the intermediate period only lasts 
for a finite  time, this average is not really well-defined
and is read off by inspection. This explains why the values
are somewhat scattered. The plot looks like the phase diagram
of a second order phase transition, with $\phi_0(0)$ as the 
``temperature'' and $\bar \phi_1$ as the order parameter.
The diagram may  be compared with similar figures in other 
work on models with spontaneous symmetry breaking, like Fig. 11 of 
Ref. \cite{Baacke:2003bt}, Figs. 6-8 in Ref. \cite{Baacke:2000fw}
and Fig. 4 in \cite{Boyanovsky:1998yp}. 
In contrast to the models studied previously, for the present model
the waterfall field always ends up at $|\phi_1| \simeq v$ as
 $ t \to \infty$, but this is a process that takes a long time.  


\subsection{The effective potential}
\label{theeffpot}
As we have noticed, the motion of the waterfall fields towards its
tree level expectation value does not seem to be driven by any strong
force. The slow and somewhat irregular motion rather suggests an 
effective potential that is quite flat, with a small inclination  
towards $\phi_1=\pm v$. Indeed it is 
well-known that for $\phi^4$ theory with $SO(N)$ symmetry and  a Mexican hat
potential the central region $|\vec\phi|<v$ becomes flat, once quantum
corrections are taken into account in the large-$N$ limit. Here a similar
phenomenon takes place. The effective potential for our 
model in the large-$N$ limit is derived in
Appendix \ref{lneffpot} and is displayed in Fig. \ref{figure:effpot}.

The region in the center is flat in the $\phi_1$-direction, it is the
region $B$ defined in Appendix \ref{lneffpot} where the gap
equation would yield the unphysical solution $\calm_a^2<0$.
At $\phi_0=0$ it is bounded by $\phi_1=\pm v$, i.e.
 by $\pm 1/\sqrt{\kappa}$ in our mass units.
 One sees that the spinodal region has disappeared and 
the effective potential has become flat in the central region. 
Of course this is the {\em equilibrium} effective potential, 
here its flatness seems to be ``realized dynamically'' by the motion of the
waterfall field, as it was found in a somewhat different 
way for the $SO(N)$ model  
in Ref. \cite{Boyanovsky:1998yp}.


\section{Summary and Conclusions}
\label{summaryandconclusions}
We have studied here a supersymmetric hybrid model in a large-$N$ 
approximation. 
The motivation for this research was the interest in the r\^ole of
quantum fluctuations in the evolution 
near and after the end of
the slow roll regime. In generalizing previous work we have taken
into account the fluctuations of all superpartners of the waterfall field. 
While the fermion field fluctuations remain small in general and
they mainly play their r\^ole in the renormalization, the
fluctuations of the scalar and pseudoscalar fields show an interesting
interplay, resulting in a peculiar behaviour of the
classical waterfall field which is different from
the behaviour found in previous analyses: 

(i) the evolution after the end of slow roll displays two stages: 
in the first (intermediate) stage the behaviour is 
similar to the one found previously for the 
hybrid model with quantum backreaction \cite{Baacke:2003bt}, 
only the scalar  field fluctuations develop and the system goes 
to a symmetric phase at
high excitations, and to a broken symmetry phase at low excitation.
In the second stage (late time) the fluctuations of the pseudoscalar
superpartner start developing and the waterfall fields moves towards its
classical vacuum expectation value. 

(ii) the dynamics of the fluctuations is, apart from an initial
spinodal evolution of the scalar field fluctuations, neither 
dominated by the spinodal instability nor by parametric resonance.
On the one hand a clean parametric resonance is suppressed
by the presence of multiple time scales, there is no simple 
periodic motion in the background fields. On the other hand
the oscillations of the effective squared masses
$\calm_a^2$ and $\calm_b^2$ around zero lead to a kind of diffusive
behaviour for the (dominant) low energy modes. 
Within short time intervals exponential and
oscillatory evolution alternate,  the fluctuations and
their integrals vary in a stochastic way. This behaviour is transmitted
to the waterfall field whose motion is an irregular drift towards
its classical vacuum expectation value. 

We think that these findings warrant a further consideration
of the hybrid model in its supersymmetric extension.
The following issues would be of interest and could be the subject
for further studies:

(i) how does the behaviour change in the presence of Hubble expansion,
i.e. in a FRW universe? This will lead to a damping of the classical
field {\em and} of the fluctuations. While the Hubble expansion is 
often neglected in studies of preheating, here, right at the end of 
inflation, it can be expected to lead to significant modifications. 
This is also the reason for which we do not want to draw premature
conclusions from our numerical results. The techniques for handling
the dynamics in an expanding universe, including renormalization,
have been developed previously \cite{Baacke:1997rs,Baacke:1999nq}.

(ii) does the peculiar dynamical behaviour lead to imprints on the
CMBR spectrum? In particular, nongaussianity may arise in such 
multifield models from the evolution near the end
of inflation, see \cite{Barnaby:2006km} and references therein. 
   
(iii) is the peculiar behaviour a genuine property of a supersymmetric
model with its specific structure of the mass terms, or is it
simply a consequence of the presence of many different time
scales, or both? 

We plan to study these questions in the near future.

\newpage

\begin{appendix}
\renewcommand{\theequation}{\Alph{section}\arabic{equation}}

\section{Renormalization}
\label{renormalization}
\setcounter{equation}{0}
The fluctuation integrals introduced in section \ref{basics} are divergent.
In dimensional regularization we find
\bea\nonumber
<a^2> &=& \calf_a= -I_{-3}(m^2)\calm_a^2 + \calf_a^{\rm fin}\kma 
\\\label{divergences}
<b^2> &=& \calf_b= -I_{-3}(m^2)\calm_b^2 + \calf_b^{\rm fin} \kma
\\\nonumber 
<\bar\psi\psi> &=& \calf_f= I_{-3}(m^2)\left[2\ddot\calm_f
+4\calm_f^3\right]+\calf_f^{\rm fin}
\pkt\eea
Here
\bea
\calf_a^{\rm fin}&=&-\frac{1}{16\pi^2}m_{a0}^2-\frac{\calm_a^2}{16\pi^2}
\ln\frac{m^2}{m_{a0}^2}+\calf_a^{\rm sub}\kma
\\
\calf_b^{\rm fin}&=&-\frac{1}{16\pi^2}m_{b0}^2-\frac{\calm_b^2}{16\pi^2}
\ln\frac{m^2}{m_{b0}^2}+\calf_b^{\rm sub}\kma
\\
\calf_f^{\rm fin}&=&\frac{1}{4\pi^2}m_{f0}^2\calm_f+\frac{1}{8\pi^2}
\ln\frac{m^2}{m_{f0}^2}\left(2\calm_f^3+\sqrt{2}\kappa\ddot\phi_0\right)
+\calf_f^{\rm sub}\kma
\eea
with
\be
\calf_{a/b}^{\rm sub}(t)=
\intk{2\omega_{a/b0}}\left[|f^{ab}_k(t)|^2+\frac{\calv_{a/b}}
{2\omega_{a/b0}^2}\right]
\ee
and \cite{Baacke:1998di}
\bea\nonumber
\calf_f^{\rm sub}(t)&=&
-2\intk{}\left[1-\frac{\omega_{f0}-m_{f0}}{\omega_{f0}}
|f_k^\psi(t)|^2-\frac{\calm_f(t)}{\omega_{f0}}\right.
\\
&&\left.+
\frac{\ddot \calm_f(t)}{4 \omega_{f0}^3}+\frac{\calm_f(t)(\calm_f^2(t)-m_{f0}^2)}
{2\omega_{f0}^3}\right]\pkt
\eea

 We introduce the renormalization constants $Z_i$ as follows:
\bea\nonumber
\phi_0 &\to& Z_0 \phi_0
\\
\phi_1 &\to& Z_1 \phi_1
\\\nonumber
\kappa &\to& Z_\kappa \kappa
\\\nonumber
v &\to& Z_1 v
\eea
The fluctuation fields $a,b,\psi$ are likewise multiplied by $Z_1$
as they belong to the same superfield as $\phi_1$, more precisely
they belong to the same $SO(N)$ multiplet of superfields as $\phi_1$.
The vacuum expectation value has been rescaled as $\phi_1$ in order
to preserve its tree level interpretation.

We first discuss the gap equation. It takes the form
\be
\calm_-^2=Z_\kappa^2Z_1^2\kappa^2\left[\phi_1^2-v^2-2I_{-3}(m^2)\calm_-^2
+\calf_-^{\rm fin}\right]
\ee 
with $\calf_-^{\rm fin}=\calf_a^{\rm fin}-\calf_b^{\rm fin}$.
This can be written as
\be
\calm_-^2(1+2\kappa^2Z_\kappa^2Z_1^2I_{-3}(m^2))=Z_\kappa^2Z_1^2\kappa^2\left[\phi_1^2-v^2
+\calf_-^{\rm fin}\right]\pkt
\ee 
In order to obtain the finite equations in the same form as the 
unrenormalized ones we have to put $1+2\kappa^2Z_\kappa^2Z_1^2I_{-3}(m^2)$ 
equal to
$Z_\kappa^2Z_1^2$
or 
\be\label{zkappaz1}
Z_\kappa^2Z_1^2=\frac{1}{1-2\kappa^2I_{-3}(m^2)} 
\ee
and obtain the finite gap equation
\be
\calm_-^2=\kappa^2\left(\phi_1^2-v^2+\calf_-^{\rm fin}\right)
\pkt\ee

Next we consider the equations of motion.
We start with the one for $\phi_1$, which after divison by 
$Z_1$ takes the form
\be
\ddot\phi_1+2Z_\kappa^2Z_0^2\kappa^2\phi_0^2\phi_1+\calm_-^2\phi_1=0
\pkt\ee
As $\calm_-^2$ is already finite we have to set
\be\label{zkappaz0}
Z_0Z_\kappa=1
\pkt\ee
This entails that also the renormalized  equation of motion for the 
field $\psi$
retains its bare form.
We now consider the equation of motion for $\phi_0$. It becomes, 
after dividing by $Z_0$ which is equivalent of multiplying with $Z_\kappa$ 
\be
\ddot \phi_0+Z_\kappa^2Z_1^2\left[2\kappa^2\phi_0\phi_1^2+
2\kappa^2\phi_0(<a^2>+<b^2>)+
\frac{1}{\sqrt{2}}\kappa<\bar\psi\psi>\right]=0
\pkt\ee
We use the relations \eqn{divergences} to rewrite this as
\bea\nonumber
&&\ddot \phi_0+Z_\kappa^2Z_1^2\left[
2\kappa^2\phi_0(\phi_1^2-4I_{-3}(m^2)\kappa^2\phi_0^2+\calf_+^{\rm fin})
\right.\\
&&\left.+\frac{1}{\sqrt{2}}\kappa I_{-3}(m^2)
(2\sqrt{2}\kappa\ddot\phi_0+8\sqrt{2}\kappa^3\phi_0^3)
+\frac{1}{\sqrt{2}}
\kappa \calf_f^{\rm fin}\right]=0
\kma\eea
where $\calf_+^{\rm fin}=\calf_a^{\rm fin}+\calf_b^{\rm fin}$.
In order to obtain again the tree level relation with finite quantities
we have to put
\be
1+2Z_\kappa^2\kappa^2Z_1^2I_{-3}(m^2)=Z_\kappa^2Z_1^2
\pkt\ee
which is the same relation we have obtained previously.

Up to now only the products $Z_0Z_\kappa$ and $Z_1Z_\kappa$ are fixed.
We now consider the renormalization of the energy.
The total energy density written in terms of the bare fields and couplings is given by
\bea\nonumber
\cale&=&\frac{1}{2}\dot \phi_0^2+\frac{1}{2}\dot \phi_1^2+
\kappa^2\phi_0^2\phi_1^2+\frac{\kappa^2}{4}\left(\phi_1^2-v^2\right)^2
\\
&&+<\frac{1}{2}\dot a^2+\frac{1}{2}k^2a^2+\frac{1}{2}\calm_a^2 a^2>
\\\nonumber
&&+<\frac{1}{2}\dot b^2+\frac{1}{2}k^2b^2+\frac{1}{2}\calm_b^2 a^2>
\\\nonumber
&&-\frac{\kappa^2}{4}\left(<a^2>-<b^2>\right)^2+ 
\frac{1}{2}\psi^\dagger\left(\vec\alpha\vec k +\beta \calm_f\right)\psi
\pkt\eea
In terms of the renormalized fields this becomes
\bea\nonumber
\cale&=&\frac{1}{2}Z_0^2\dot \phi_0^2+\frac{1}{2}Z_1^2\dot \phi_1^2+
Z_0^2Z_1^2Z_\kappa^2\kappa^2\phi_0^2\phi_1^2+
Z_1^4Z_\kappa^2\frac{\kappa^2}{4}\left(\phi_1^2-v^2\right)^2
\\\label{erenorm1}
&&+Z_1^2<\frac{1}{2}\dot a^2+\frac{1}{2}k^2a^2+\frac{1}{2}\calm_a^2 a^2>
\\\nonumber
&&+Z_1^2<\frac{1}{2}\dot b^2+\frac{1}{2}k^2b^2+\frac{1}{2}\calm_b^2 a^2>
\\\nonumber
&&-Z_\kappa^2Z_1^4\frac{\kappa^2}{4}\left(<a^2>-<b^2>\right)^2+ 
\frac{1}{2}Z_1^2\psi^\dagger\left(\vec\alpha\vec k +\beta \calm_f\right)\psi
\pkt\eea

Separating the divergent parts we have
\bea\nonumber
\cale_{\rm fl,a}&=&<\frac{1}{2}\dot a^2+\frac{1}{2}k^2a^2+\calm_a^2 a^2>
\\&=& -\frac{1}{4}I_{-3}(m^2)\calm_a^4+\cale_{\rm fl,a}^{\rm fin}\kma
\\\nonumber
\cale_{\rm fl,b}&=&<\frac{1}{2}\dot b^2+\frac{1}{2}k^2b^2+\calm_b^2 b^2>
\\&=& -\frac{1}{4}I_{-3}(m^2)\calm_b^4+\cale_{\rm fl,b}^{\rm fin}\kma
\\\nonumber
\cale_{\rm fl,f}&=&\frac{1}{2}\bar\psi\left(\vec\alpha\vec k +\beta \calm_f\right)\psi
\\&=&\frac{1}{2}I_{-3}(m^2)\left[2\kappa^2\dot\phi_0^2+
4\kappa^4\phi_0^4\right] + \cale_{\rm fl,f}^{\rm fin}
\pkt\eea
Here 
\bea 
\cale_{\rm fl,a}^{\rm fin}&=&\frac{1}{128\pi^2}m_{a0}^4-
\frac{1}{32\pi^2}m_{a0}^2\calm_a^2-\frac{1}{64\pi^2}\ln\frac{m^2}{m_{a0}^2}
\calm_a^4
+\cale_{\rm fl,a}^{\rm sub}\kma
\\
\cale_{\rm fl,b}^{\rm fin}&=&\frac{1}{128\pi^2}m_{b0}^4-
\frac{1}{32\pi^2}m_{b0}^2\calm_b^2-\frac{1}{64\pi^2}\ln\frac{m^2}{m_{b0}^2}
\calm_b^4+\cale_{\rm fl,b}^{\rm sub}\kma
\\\nonumber
\cale_{\rm fl,f}^{\rm fin}&=&-\frac{1}{64\pi^2}m_{f0}^4+
\frac{1}{16\pi^2}m_{f0}^2\calm_f^2+\frac{1}{32\pi^2}\ln\frac{m^2}{m_{f0}^2}
(\calm_f^4+\dot\calm_f^2)
\\&&+\cale_{\rm fl,f}^{\rm sub}\kma
\eea
where the superscript 'sub' refers to the subtracted fluctuation integrals
which can be found in detailed form  in Refs. \cite{Baacke:1998di,Baacke:1996se}.

 The divergent parts of the three fluctuation energies combine into
\bea\nonumber
\cale_{\rm  fl}^{\rm div}&=&
-\frac{1}{4}I_{-3}(m^2)\left[\calm_a^4+\calm_b^4-2\calm_f^4\right]
+I_{-3}(m^2)\kappa^2\dot\phi_0^2
\\
&=&-\frac{1}{2}I_{-3}(m^2)\calm_-^4
+I_{-3}(m^2)\kappa^2\dot\phi_0^2
\pkt\eea
In the total energy the fluctuations contribute a term
\be
-\frac{Z_1^2}{2}I_{-3}(m^2)\calm_-^4
+Z_1^2\cale_{\rm fl}^{\rm fin}+Z_1^2I_{-3}(m^2)\kappa^2\dot\phi_0^2
\pkt\ee
If we want the energy density to keep its bare form in 
terms of finite quantities 
we have to set $Z_1=1$. Then in Eq. \eqn{erenorm1} the second and third term
have already their bare form in terms of renormalized quantities.
The first term and $\dot \phi_0^2$-term in the divergent fluctuation
energies combine into
\be
\frac{1}{2}\left(Z_0^2+2\kappa^2I_{-3}(m^2)\right)\dot \phi_0^2
\pkt\ee
This takes the canonical form if
\be
Z_0^2=1-2\kappa^2I_{-3}(m^2)
\pkt\ee
This is consistent with Eq \eqn{zkappaz0} if
\be
Z_\kappa^2=Z_0^{-2}=\frac{1}{1-2\kappa^2I_{-3}(m^2)}
\ee
and this is also consistent with Eq. \eqn{zkappaz1} if $Z_1=1$ as obtained
previously.
 
The fourth term in the energy density combines with the seventh term into
\bea\nonumber
&&Z_\kappa^2\frac{\kappa^2}{4}(\phi_1^2-v^2)^2
-Z_\kappa^2\frac{\kappa^2}{4}\left(<a^2>-<b^2>\right)^2
\\\nonumber
&&=\frac{1}{4}\calm_-^2\left[\phi_1^2-v^2-(<a>^2-<b>^2)\right]
\\
&&=\frac{1}{2}\calm_-^2\left(\phi_1^2-v^2\right)-\frac{1}{4Z_\kappa^2\kappa^2}
\calm_-^4
\\\nonumber
&&=\frac{1}{2}\calm_-^2\left(\phi_1^2-v^2\right)-\frac{1}{4\kappa^2}
\calm_-^4 +\frac{1}{2}I_{-3}\calm_-^4
\pkt\eea
The last term in the last equation cancels with the $\calm_-^4$-term
in the fluctuation energies. So with the choices of the $Z_i$ given above
the energy density and the equations of motion are finite.
The energy density now becomes
\be
\cale=\frac{1}{2}\dot \phi_0^2+\frac{1}{2}\dot \phi_1^2+
\kappa^2\phi_0^2\phi_1^2 +\cale_{\rm fl}^{\rm fin}
+\frac{1}{2}\calm_-^2\left(\phi_1^2-v^2\right)-\frac{1}{4\kappa^2}\calm_-^4
\pkt\ee
Here the gap equation has been implemented. Indeed,
variation with respect to $\calm_-^2$ yields the gap equation
\be
\calm_-^2(t)=\kappa^2\left(\phi_1^2(t)-v^2+
2\frac{\delta\cale_{\rm fl}^{\rm fin}}
{\delta \calm_-^2(t)}\right)
\pkt\ee
The equations of motion for the fluctuations as well as the one for
$\phi_1$ have exactly the same form as the bare equations if we use
the masses $\calm_a^2,\calm_b^2,\calm_f^2$ and $\calm_-^2$.
The equation of motion for $\phi_0$ has the form
\be
\ddot\phi_0+2\kappa^2\phi_0\left(\phi_1^2+\calf_+^{\rm fin}\right)+
\frac{\kappa}{\sqrt{2}}\calf_f^{\rm fin}=0
\pkt\ee
This equation of motion cannot be solved trivially with respect to 
$\ddot\phi_0$ as $\calf_f^{\rm fin}$ explicitly contains $\ddot\phi_0$.
So the term $\ddot\phi_0$ appears with a factor
\be
\calc_0=1+2\frac{\kappa^2}{16\pi^2}\ln\frac{m^2}{m_{f0}^2}
\pkt\ee
Likewise the finite gap equation contains $\calm_-^2$ on both sides.
At $t=0$ this is not a problem as it is solved by iteration anyway.
But if we want to express $\calm_-^2$ at finite $t$ by the subtracted
fluctuation integrals we have to write it in a modified form.
We have 
\bea\nonumber
\calm_-^2&=&\kappa^2\left[\phi_1^2-v^2-\frac{1}{16\pi^2}(m_{a0}^2-m_{b0}^2)
 +\frac{\calm_f^2}{16\pi^2}\ln\frac{m^2_{a0}}{m^2_{b0}}\right.
\\&&\left. +
\frac{\calm_-^2}{16\pi^2}\ln\frac{m^2_{a0}m^2_{b0}}{m^4}+
\calf_a^{\rm sub}-\calf_b^{\rm sub}\right]
\kma\eea
or 
\be
\calm_-^2=\calc_-\kappa^2\left[\phi_1^2-v^2-
\frac{1}{16\pi^2}(m_{a0}^2-m_{b0}^2)
 +\frac{\calm_f^2}{16\pi^2}\ln\frac{m^2_{a0}}{m^2_{b0}}
+\calf_a^{\rm sub}-\calf_b^{\rm sub}\right]
\kma\ee
with
\be
\calc_-=\frac{1}{\displaystyle
1-2 \frac{\kappa^2}{16\pi^2}\ln\frac{m_{a0}m_{b0}}{m^2}}
\pkt\ee


\section{The large-N effective potential}
\label{lneffpot}

\setcounter{equation}{0}
The energy density for constant fields is the effective potential.
This effective potential is obtained 
in the $2PI$ formalism by maximizing with respect to 
$\calm_-^2$ the action as a functional of the fields and masses. In our
model and in the large-$N$ approximation
this action takes the form
\bea\nonumber
\calv(\calm_-^2,\phi_0,\phi_1)&=&
\kappa^2\phi_0^2\phi_1^2+\frac{1}{2}\calm_-^2(\phi_1^2-v^2)-
\frac{1}{4\kappa^2}\calm_-^4
\\\nonumber
&&-\frac{\calm_a^4}{128\pi^2}\left(3+2\ln\frac{m^2}{\calm_a^2}\right)
-\frac{\calm_b^4}{128\pi^2}\left(3+2\ln\frac{m^2}{\calm_b^2}\right)
\\\nonumber
&&+\frac{\calm_f^4}{64\pi^2}\left(3+2\ln\frac{m^2}{\calm_f^2}\right)
\pkt\eea
Variation with respect to the mass $\calm_-^2$
 yields the gap equation:
\bea
\frac{\partial \calv(\calm_-^2,\phi_0,\phi_1)}{\partial\calm_-^2}
&=&\frac{1}{2}(\phi_1^2-v^2)-\frac{1}{2\kappa^2}\calm_-^2
\\\nonumber
&&
-\frac{\calm_a^2}{32\pi^2}\left(1+\ln\frac{m^2}{\calm_a^2}\right)
+\frac{\calm_b^2}{32\pi^2}\left(1+\ln\frac{m^2}{\calm_b^2}\right)=0
\kma\eea
or
\be
\calm_-^2=\kappa^2\left[\phi_1^2-v^2-\frac{\calm_a^2}{16\pi^2}
\left(1+\ln\frac{m^2}{\calm_a^2}\right)+
\frac{\calm_b^2}{16\pi^2}\left(1+\ln\frac{m^2}{\calm_b^2}\right)\right]
\pkt\ee
The solution of this equation is to be inserted into $\calv(\calm_-^2,\phi_0,
\phi_1)$ in order to obtain $V_{\rm eff}(\phi_0,\phi_1)$.

The gap equation has a solution only in a restricted part of the
$\phi_0,\phi_1$ plane. Both $\calm_a^2$ and $\calm_b^2$ have 
to be positive, and 
such a solution does not exist everywhere. The boundaries of this region 
are traced by the conditions $\calm_a^2=0$ and $\calm_b^2=0$. Outside
this region the maximum of the variational potential is to be taken
{\em at the boundaries}. In spontaneously broken $\phi^4$ theory 
at large $N$
this construction leads to a flat potential in the central region $|\vec \phi|<v$
of the ``Mexican hat'' (see related discussions in  Ref.
\cite{Bardeen:1983st}).

Here the boundaries of the regions where the gap equation
yields the unphysical solutions $\calm_b^2<0$
(region A) and $\calm_a^2<0$ (region B) are determined by the conditions
\bea
(\partial A)\;\;&& \calm_b^2=0\;\;;\;\;  \calm_-^2=2\kappa^2\phi_0^2\;\;;\;\; 
\calm_a^2=4\kappa^2\phi_0^2\kma\\
(\partial B)\;\;&& \calm_a^2=0\;\;;\;\;  \calm_-^2=-2\kappa^2\phi_0^2\;\;;\;\; 
\calm_b^2=4\kappa^2\phi_0^2\pkt
\eea

In the regions $A$ and $B$ the masses $\calm_a^2$ and $\calm_b^2$,
respectively, retain their boundary values, i.e. zero.
Then the effective potential can be given analytically
as
\bea\label{VeffA}
V^A_{\rm eff}&=&\kappa^2\phi_0^2\phi_1^2+\kappa^2\phi_0^2
(\phi_1^2-v^2)-
\kappa^2\phi_0^4
\\\nonumber
&&-\frac{16\kappa^4\phi_0^4}{128\pi^2}
\left(3+2\ln\frac{m^2}{4\kappa^2\phi_0^2}\right)
\\
&&+\frac{4\kappa^4\phi_0^4}{64\pi^2}
\left(3+2\ln\frac{m^2}{2\kappa^2\phi_0^2}\right)
\eea
and
\bea\label{VeffB}
V^B_{\rm eff}&=&\kappa^2\phi_0^2\phi_1^2-\kappa^2\phi_0^2
(\phi_1^2-v^2)-
\kappa^2\phi_0^4
\\\nonumber
&&-\frac{16\kappa^4\phi_0^4}{128\pi^2}
\left(3+2\ln\frac{m^2}{4\kappa^2\phi_0^2}\right)
\\
&&+\frac{4\kappa^4\phi_0^4}{64\pi^2}
\left(3+2\ln\frac{m^2}{2\kappa^2\phi_0^2}\right)
\eea
The latter potential is independent of $\phi_1$ and replaces the 
potential in and around the spinodal region.

The conditions $(\partial A)$ and $(\partial B)$ do not determine the regions
of validity in terms of the fields $\phi_0$ and $\phi_1$.
A way of obtaining these regions in a numerical code is to solve
the gap equation admitting negative values of $\calm_a^2$ and
$\calm_b^2$, taking the absolute values in the logs. If
$\calm_a^2$ is found to be negative one is in region $B$, if
$\calm_b^2$ is found to be negative one is in region $A$.
As long as the potential $\calv$ has just one maximum as a
function of $\calm_-^2$ this recipe is rigorous.

If one omits the fluctuation terms, the  region where the
potential $V^B_{\rm eff}$ is independent of $\phi_1$ is the region inside
$|\phi_1|< \sqrt{v^2-2\phi_0^2}$ which includes the spinodal region.
\end{appendix}

\bibliography{lnshy}
\bibliographystyle{h-physrev4.bst}

\newpage

\begin{figure}[htb]
\begin{center}
\includegraphics[scale=0.35]{masszbs.eps}
\vspace*{6mm}
\caption{\label{figure:masszbs} Evolution at the  transition
between the  slow roll and  the intermediate period, 
for $\kappa=0.1$ and $\phi_0(0)=4$.}
\end{center}

\begin{center}
\vspace*{6mm}
\includegraphics[scale=0.35]{masszbl.eps}
\vspace*{6mm}
\caption{\label{figure:masszbl} Evolution at the  transition
between the  slow roll and  the intermediate period, 
for $\kappa=0.1$ and $\phi_0(0)=64$ .}
\end{center}
\end{figure}

\begin{figure}[htb]
\begin{center}
\includegraphics[scale=0.4]{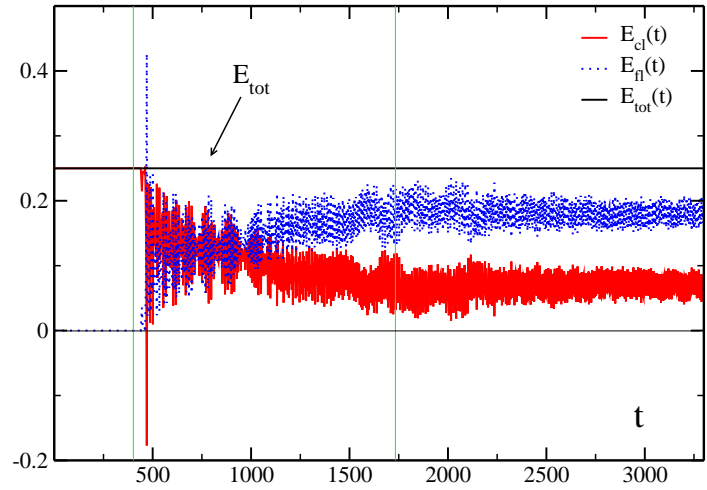}
\vspace*{6mm}
\caption{\label{figure:energysm} Evolution of the  classical
and fluctuation energies after the slow roll period.
for $\kappa=0.1$ and $\phi_0(0)=4$.}
\end{center}
\end{figure} 

\begin{figure}[htb]
\begin{center}
\includegraphics[scale=0.4]{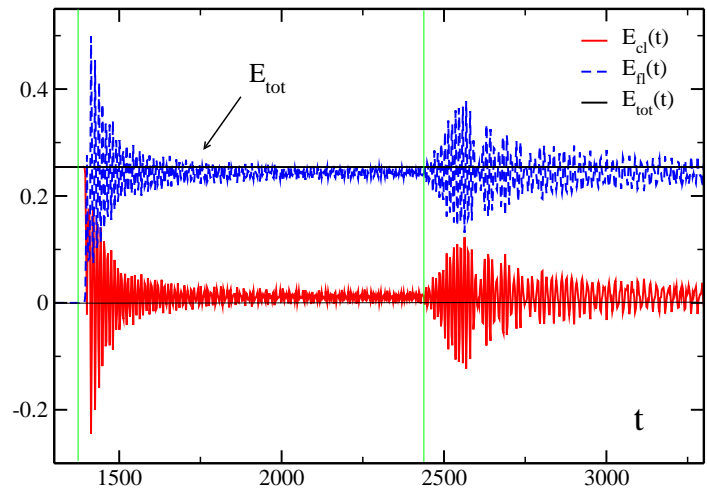}
\vspace*{6mm}
\caption{\label{figure:energylm} Evolution at the  classical
and fluctuation energy densities after slow period. 
for $\kappa=0.1$ and $\phi_0(0)=64$.}
\end{center}
\end{figure} 

\begin{figure}[htb]
\begin{center}
\includegraphics[scale=0.3]{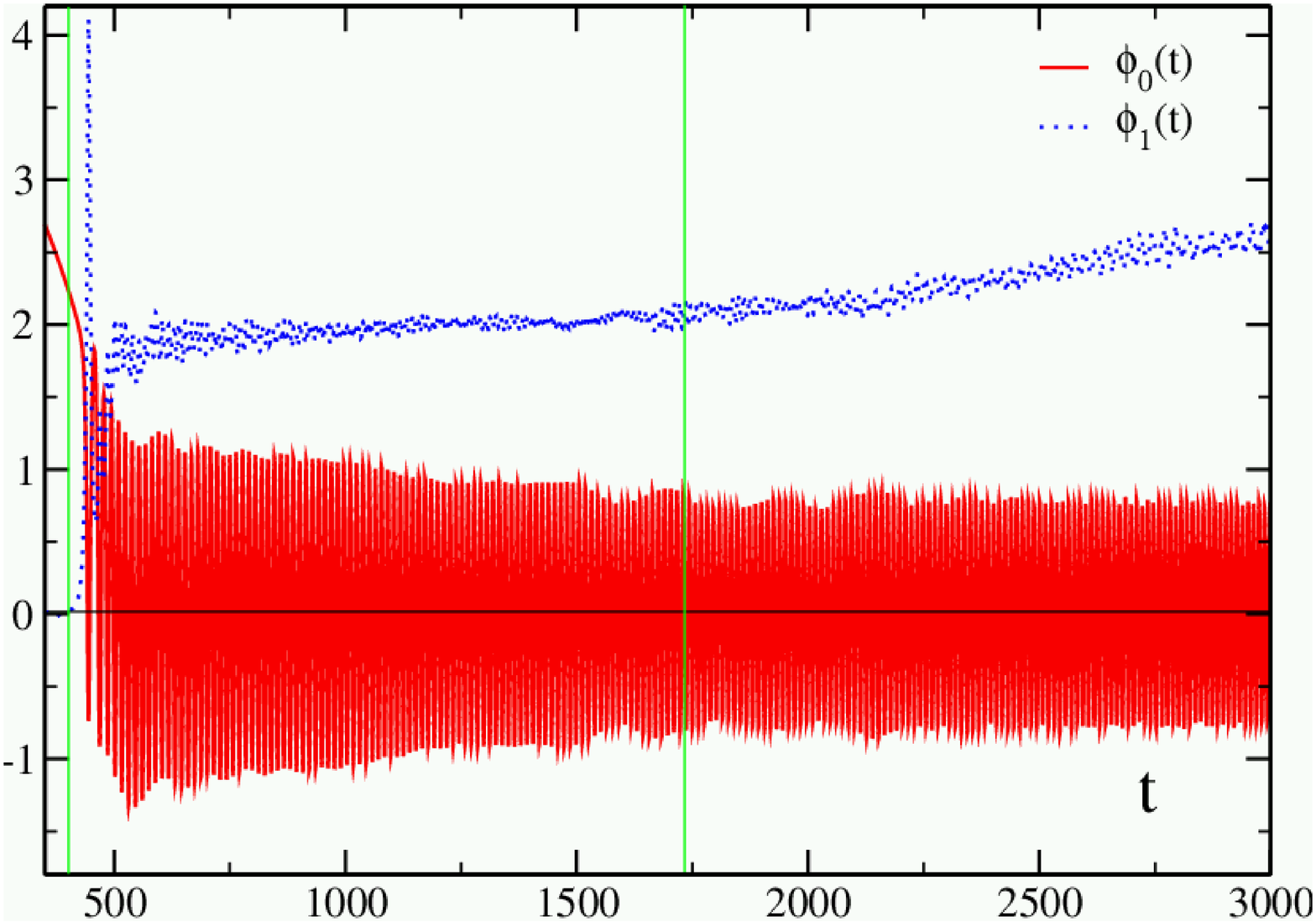}
\vspace*{6mm}
\caption{\label{figure:phism2} Evolution of the
classical fields during the intermediate period 
for $\kappa=0.1$ and $\phi_0(0)=4$.}
\end{center}
\end{figure} 

\begin{figure}[htb]
\begin{center}
\includegraphics[scale=0.3]{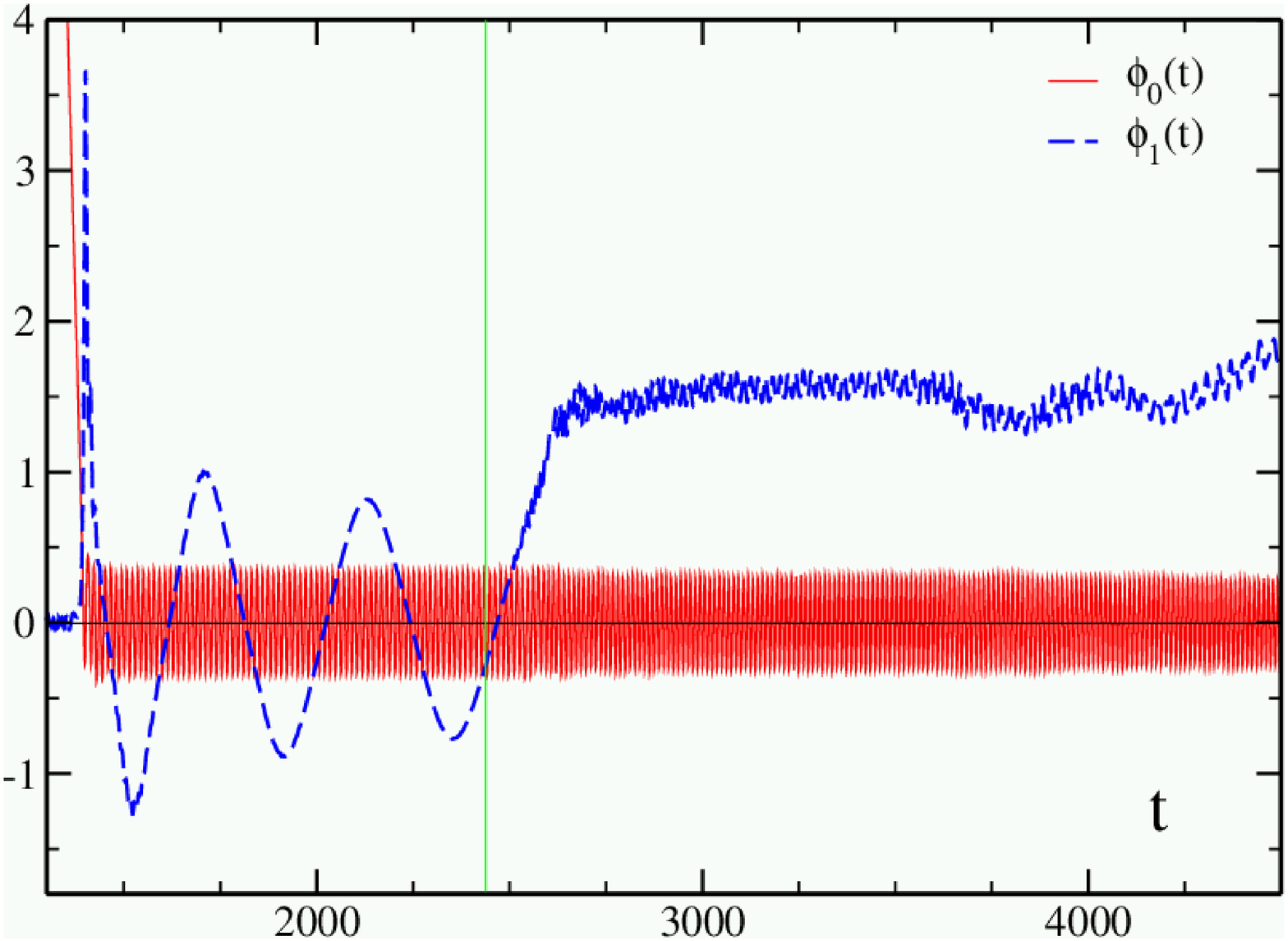}
\vspace*{6mm}
\caption{\label{figure:philm2} Evolution of the 
classical fields during  the intermediate period, 
for $\kappa=0.1$ and $\phi_0(0)=64$.}
\end{center}
\end{figure} 

\begin{figure}[htb]
\begin{center}
\includegraphics[scale=0.3]{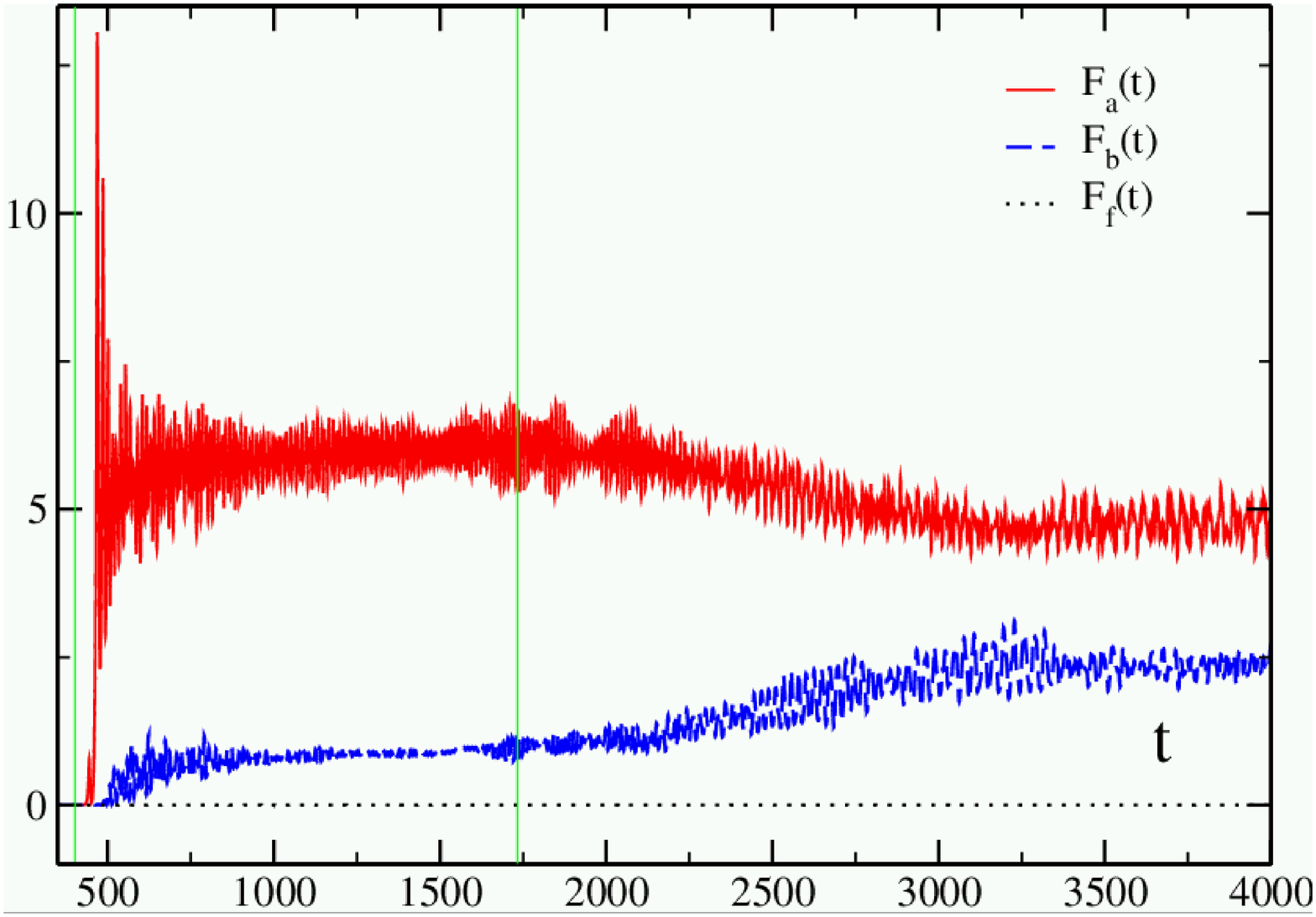}
\vspace*{6mm}
\caption{\label{figure:fluctsm2} Evolution of the
classical fields during the intermediate period 
for $\kappa=0.1$ and $\phi_0(0)=4$.}
\end{center}
\end{figure} 

\begin{figure}[htb]
\begin{center}
\includegraphics[scale=0.3]{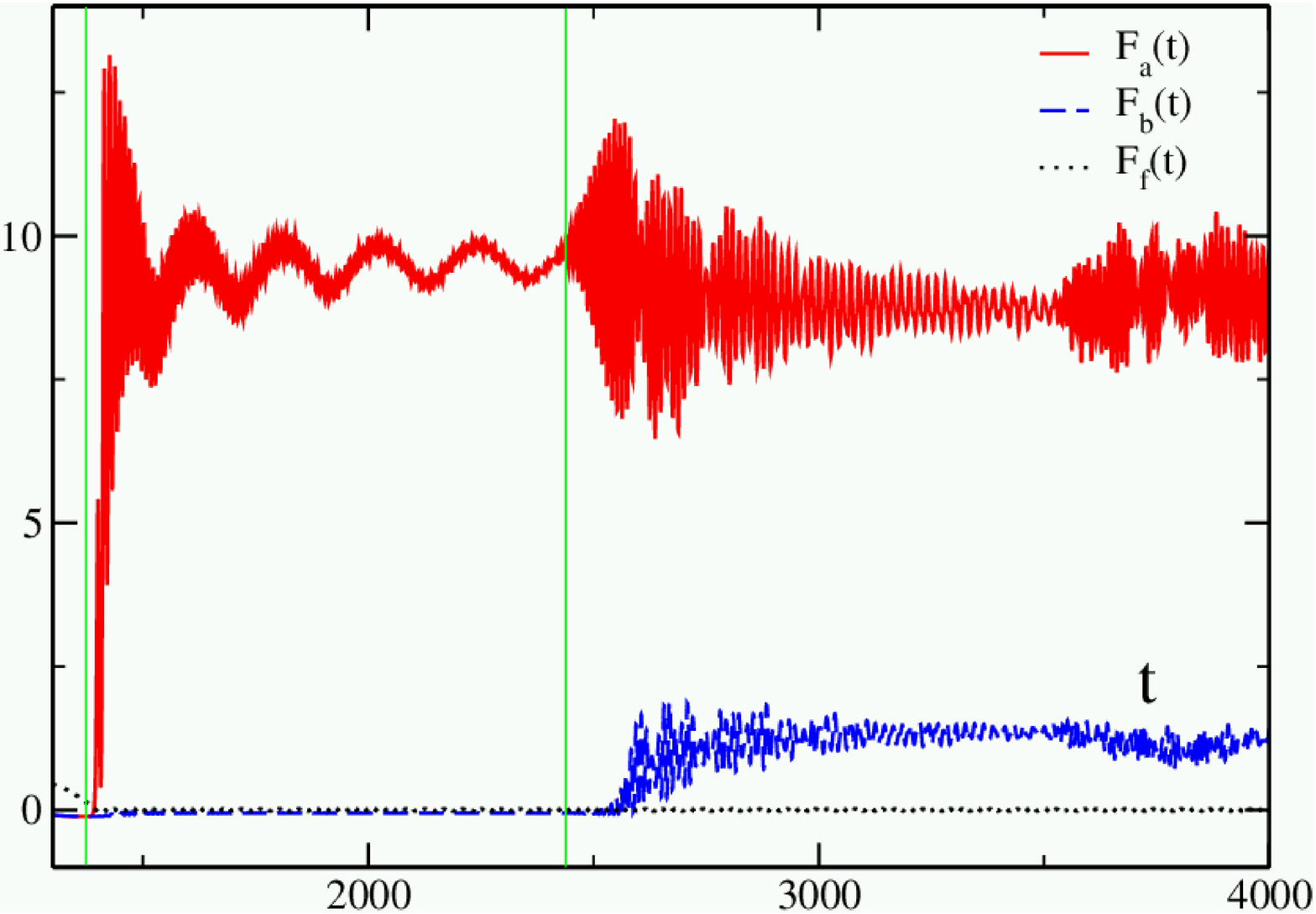}
\vspace*{6mm}
\caption{\label{figure:fluctlm2} Evolution of the 
classical fields during  the intermediate period, 
for $\kappa=0.1$ and $\phi_0(0)=64$.}
\end{center}
\end{figure} 
 
\begin{figure}[htb]
\begin{center}
\includegraphics[scale=0.25]{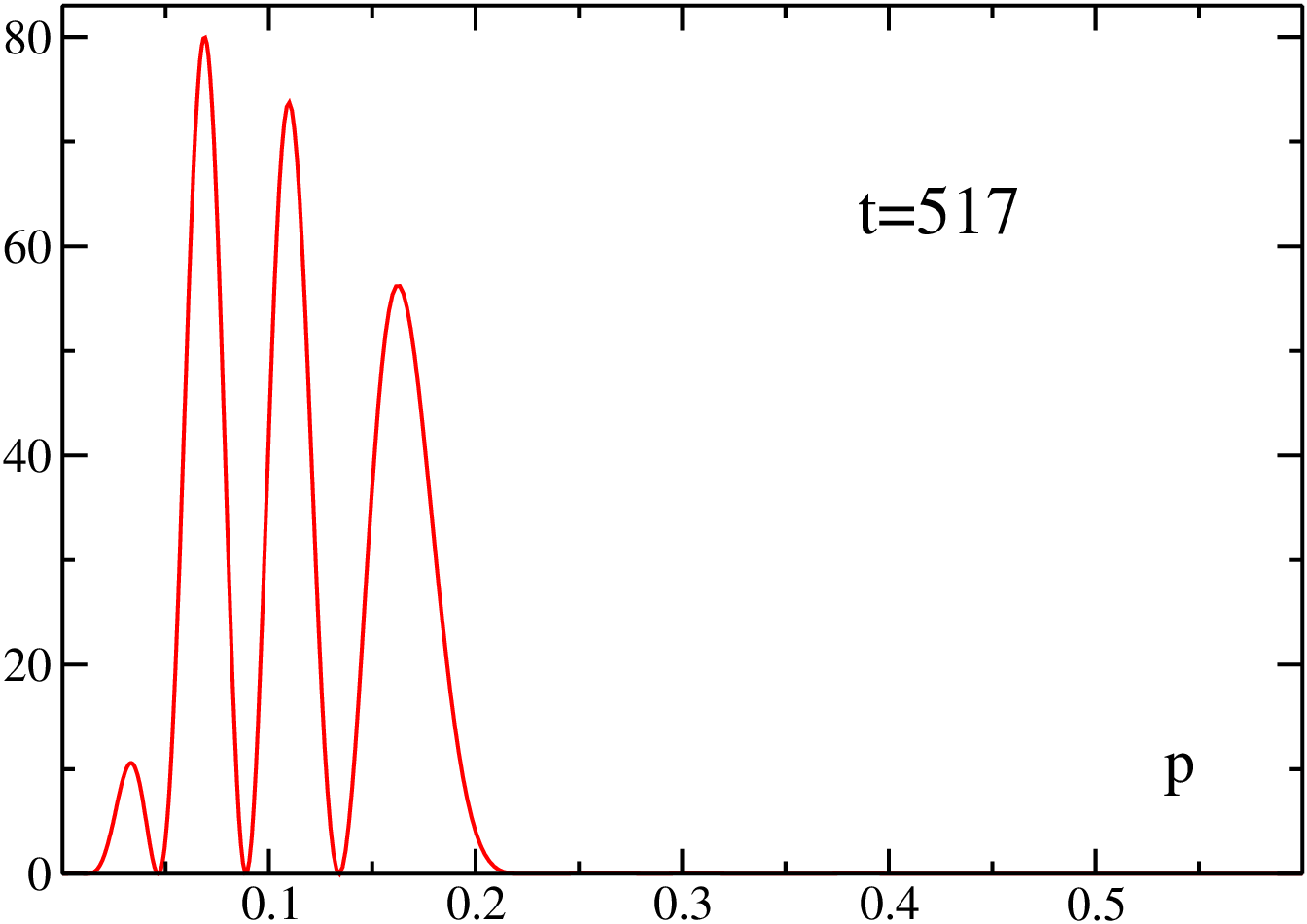}~~~~
\includegraphics[scale=0.25]{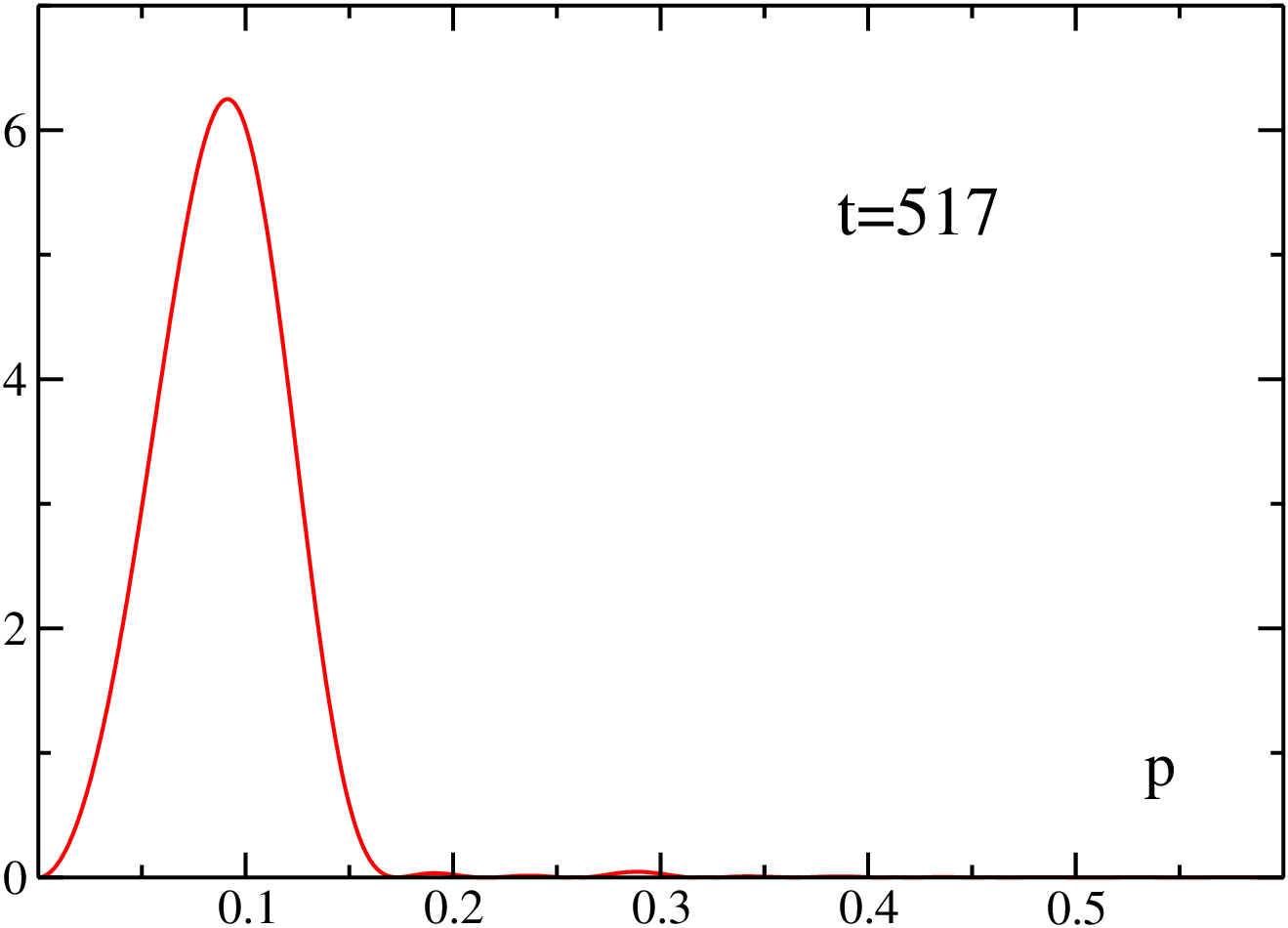}
\vspace*{6mm}
\caption{\label{figure:spectrsaba} Momentum spectra for the fluctuations
of the fields $a$ (left) and $b$ (right) right after slow roll, 
for $\kappa=0.1$ and $\phi_0(0)=4$.}
\end{center}
\end{figure} ~~
\begin{figure}[htb]
\begin{center}
\includegraphics[scale=0.25]{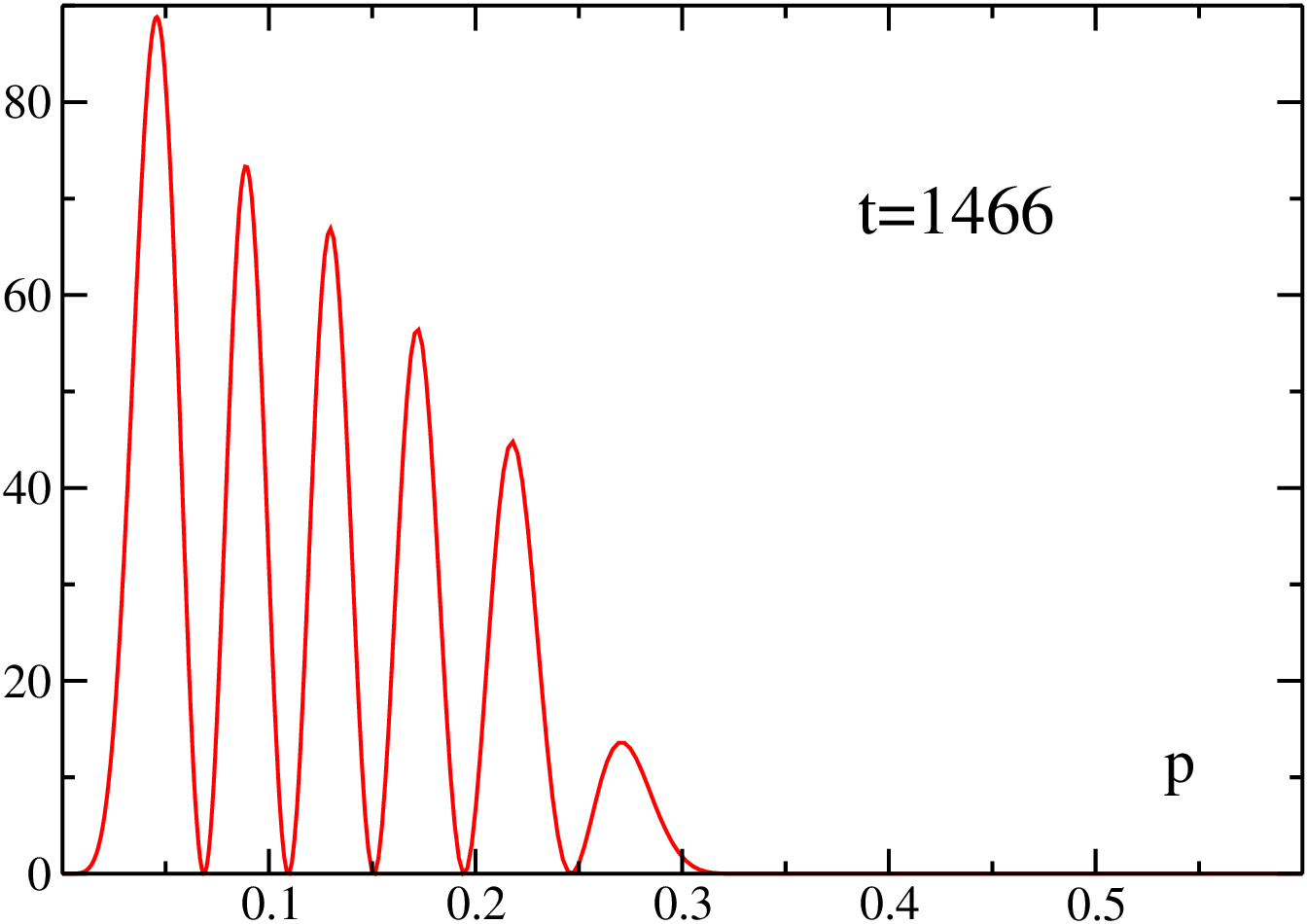}~~~~
\includegraphics[scale=0.25]{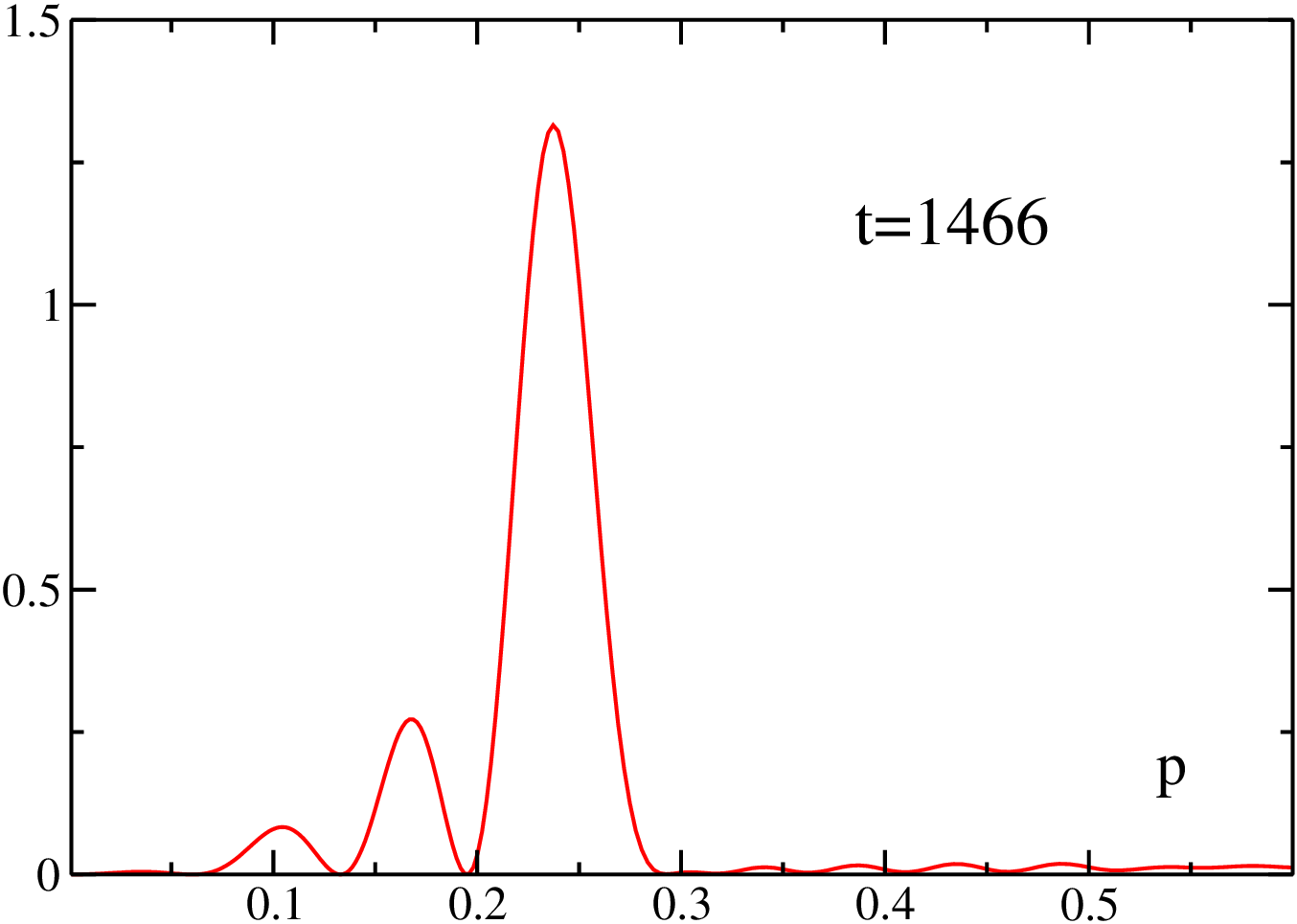}
\caption{\label{figure:spectrlaba} Momentum spectra for the fluctuations
of the fields $a$ (left) and $b$ (right) right after slow roll, 
for $\kappa=0.1$ and $\phi_0(0)=64$.}
\end{center}
\end{figure} 

\begin{figure}[htb]
\begin{center}
\includegraphics[scale=0.25]{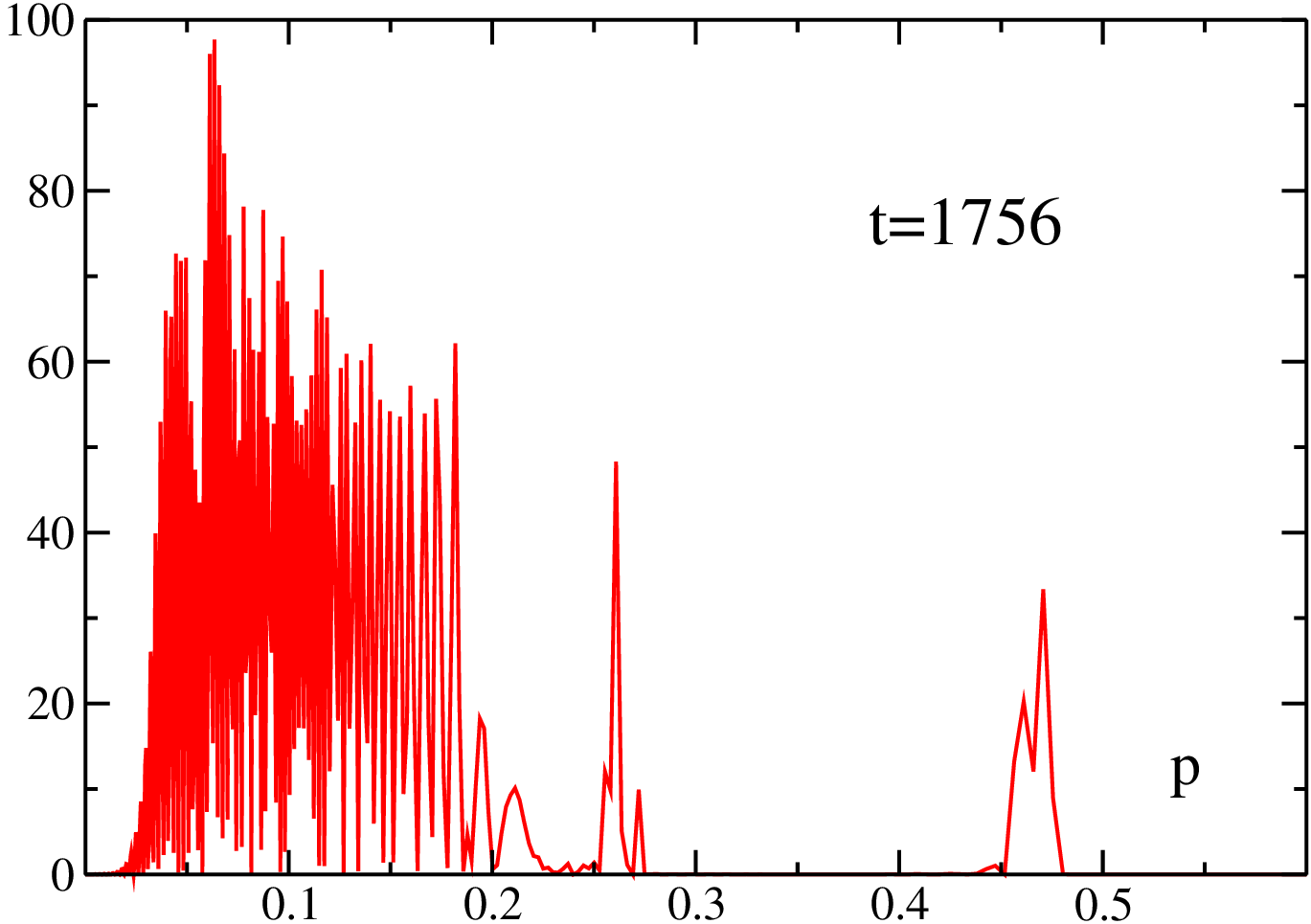}~~~~
\includegraphics[scale=0.25]{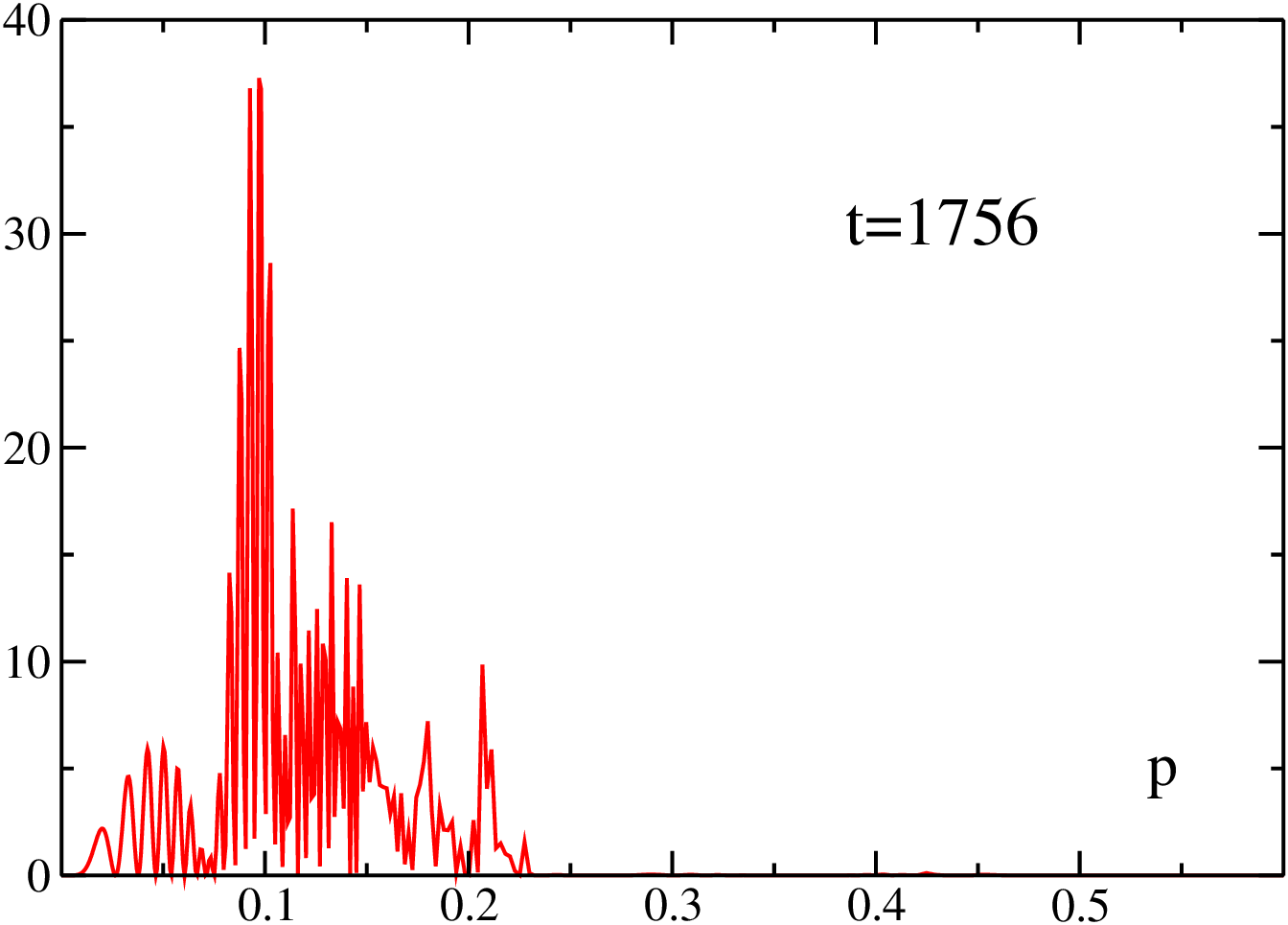}
\caption{\label{figure:spectrsabm} Momentum spectra for the fluctuations
of the fields $a$ (left) and $b$ (right)  after the intermediate period, 
for $\kappa=0.1$ and $\phi_0(0)=4$.}
\end{center}
\end{figure} ~~
\begin{figure}[htb]
\begin{center}
\includegraphics[scale=0.25]{spectrlam.eps}~~~~
\includegraphics[scale=0.25]{spectrlbm.eps}
\caption{\label{figure:spectrlabm} Momentum spectra for the fluctuations
of the fields $a$ (left) and $b$ (right) right after the intermediate period, 
for $\kappa=0.1$ and $\phi_0(0)=64$.}
\end{center}
\end{figure} 

\begin{figure}[htb]
\begin{center}
\includegraphics[scale=0.4]{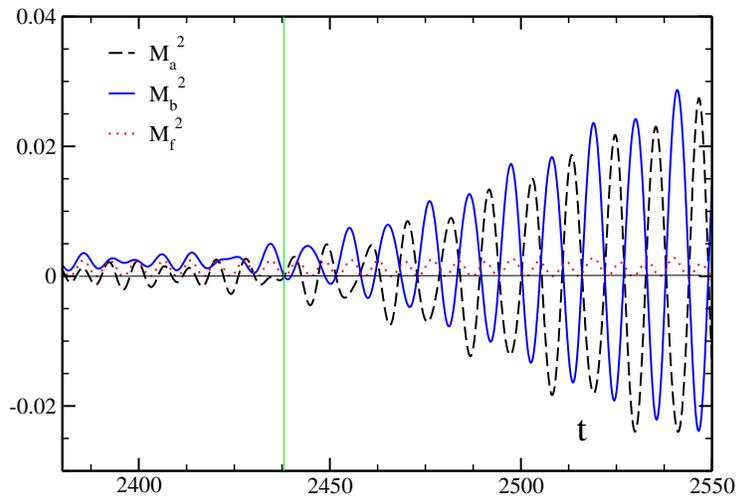}
\vspace*{6mm}
\caption{\label{figure:masszel} Evolution of the 
masses at the transition between intermediate and late time
regime, for $\kappa=0.1$ and $\phi_0(0)=4$.}
\end{center}
\end{figure} 

\begin{figure}[htb]
\begin{center}
\includegraphics[scale=0.3]{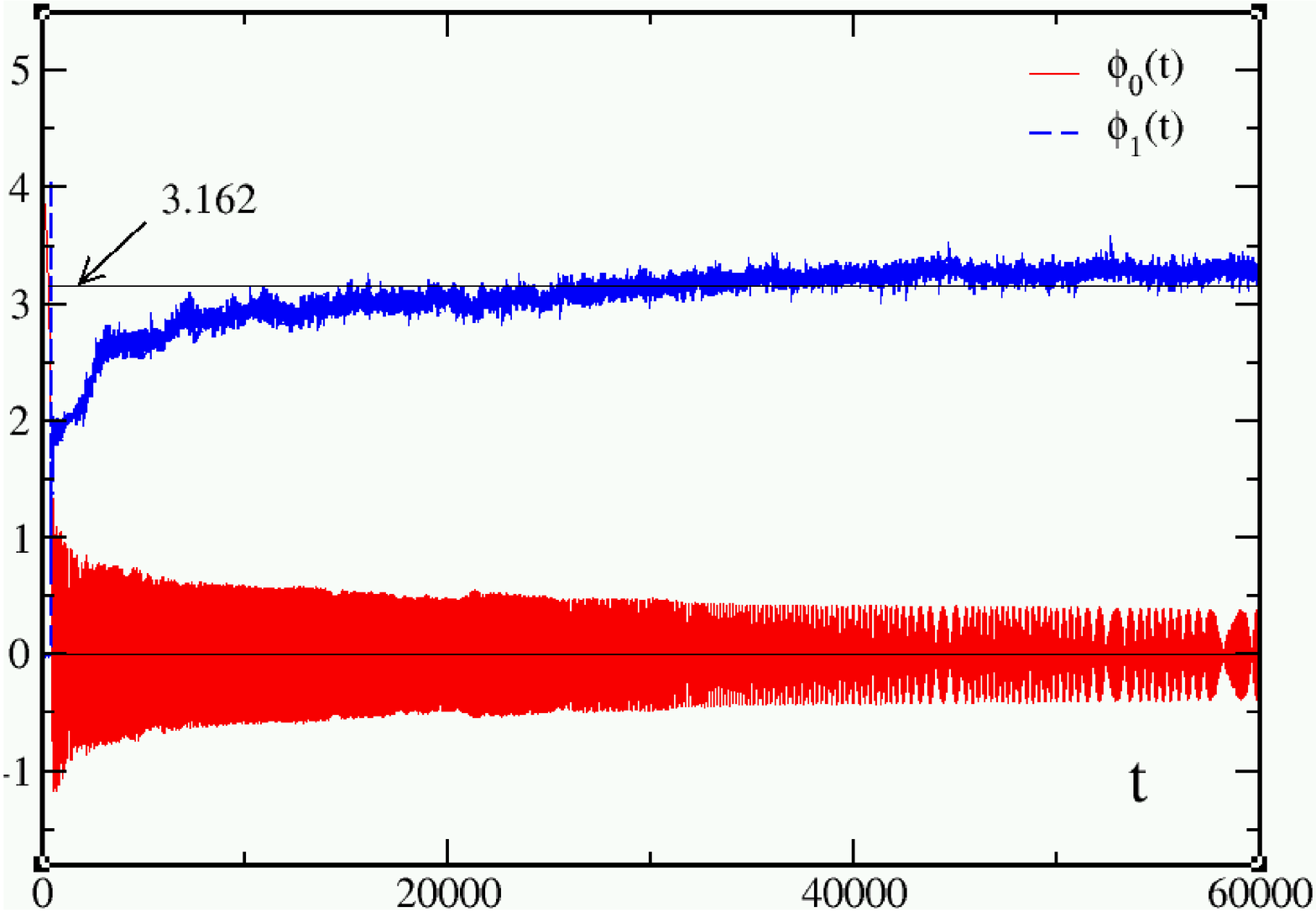}
\vspace*{6mm}
\caption{\label{figure:phisl} Evolution of the
classical fields during the intermediate period 
for $\kappa=0.1$ and $\phi_0(0)=4$.}
\end{center}
\end{figure}

\begin{figure}[htb]
\begin{center}
\includegraphics[scale=0.3]{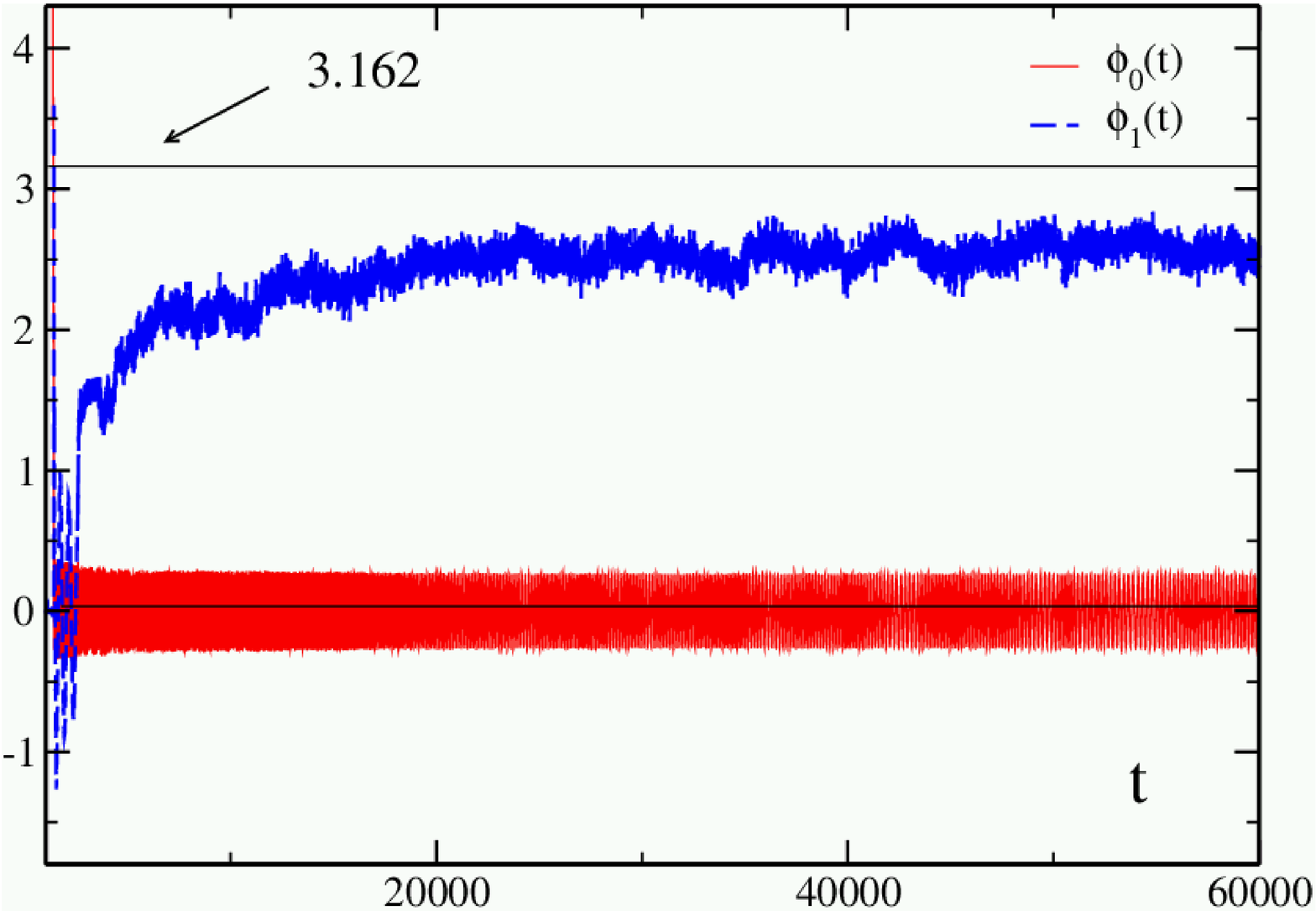}
\vspace*{6mm}
\caption{\label{figure:phill} Evolution of the 
classical fields during  the intermediate period, 
for $\kappa=0.1$ and $\phi_0(0)=64$.}
\end{center}
\end{figure}

\begin{figure}[htb]
\begin{center}
\includegraphics[scale=0.25]{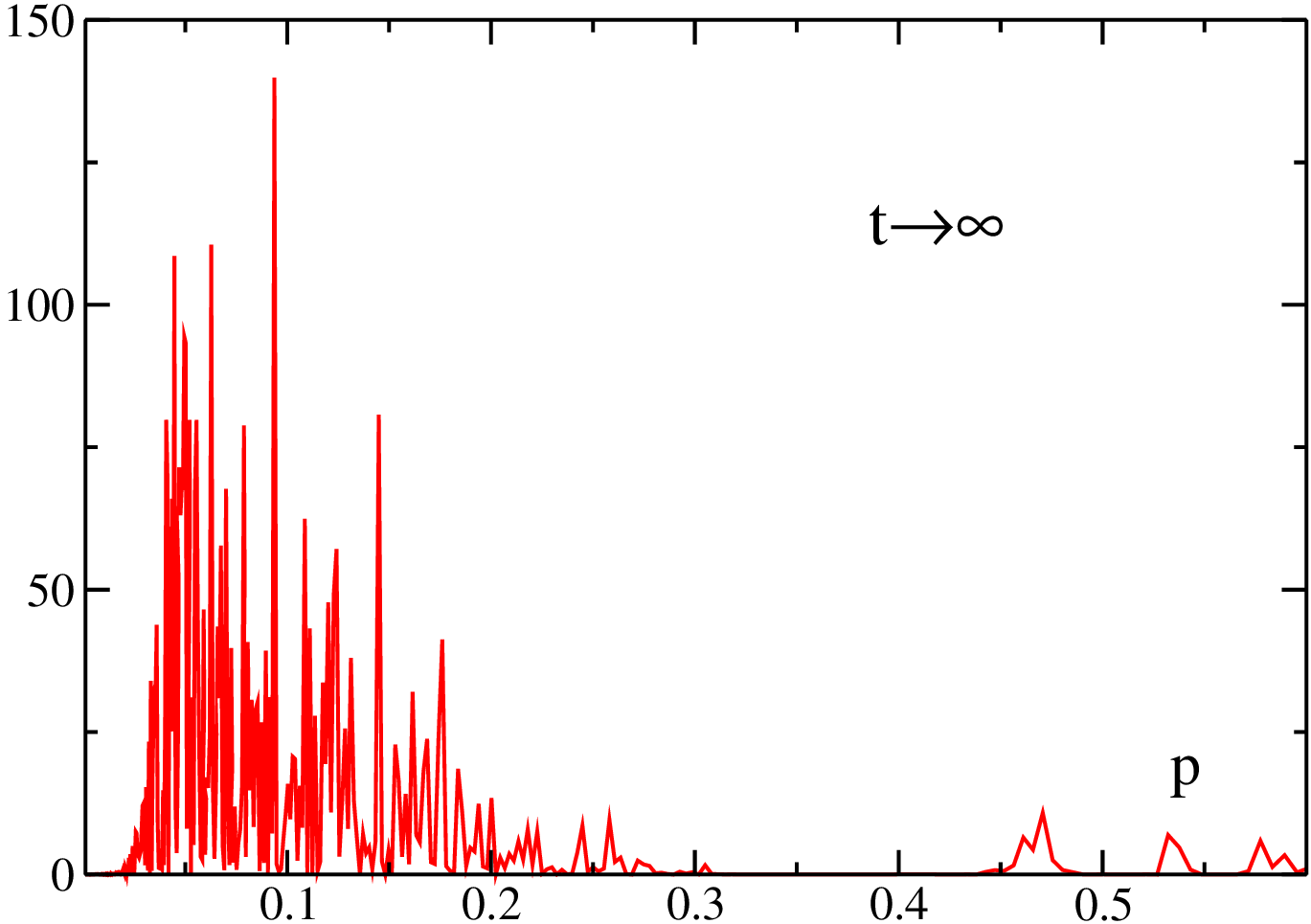}~~~
\includegraphics[scale=0.25]{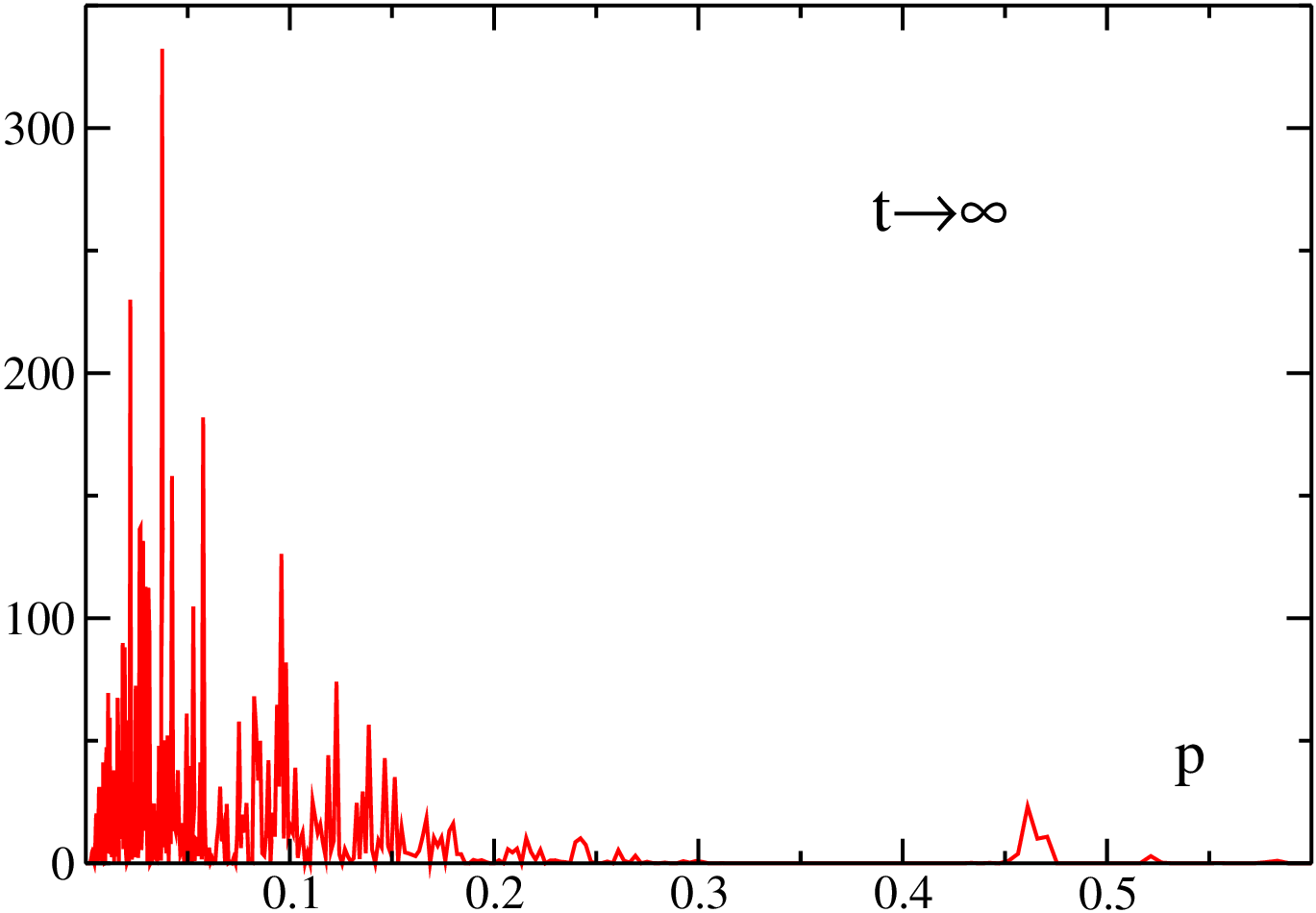}
\caption{\label{figure:spectrsabl} Momentum spectra for the fluctuations
of the fields $a$ (left) and $b$ (right) at late times, 
for $\kappa=0.1$ and $\phi_0(0)=4$.}
\end{center}
\end{figure} ~~
\begin{figure}[htb]

\begin{center}
\includegraphics[scale=0.25]{spectrlal.eps}~~~
\includegraphics[scale=0.25]{spectrlbl.eps}
\caption{\label{figure:spectrlabl} Momentum spectra for the fluctuations
of the fields $a$ (left) and $b$ (right) at late times, 
for $\kappa=0.1$ and $\phi_0(0)=64$.}
\end{center}
\end{figure} 

\begin{figure}
\begin{center}
\includegraphics[scale=0.8]{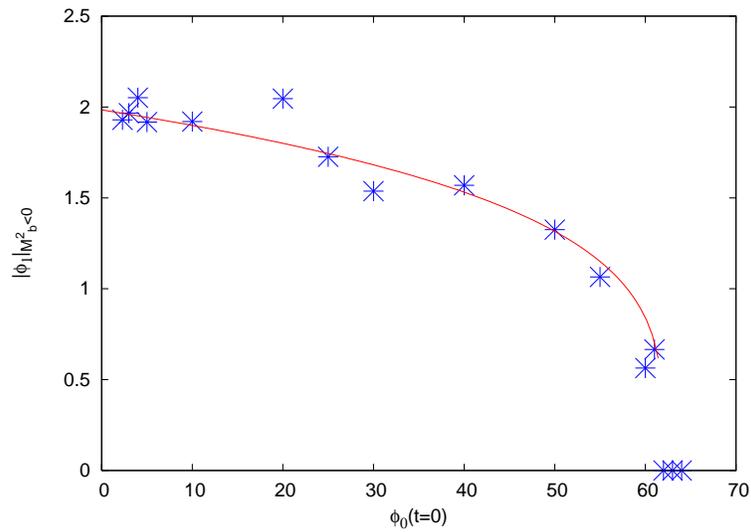}
\vspace*{6mm}
\caption{\label{figure:phasetrans} Averages of the waterfall field
in the intermediate period as function of the initial value $\phi_0(0)$,
for  $\kappa=0.1$.}
\end{center}
\end{figure} 
\newpage
\begin{figure}[t]
\begin{center}
\includegraphics[scale=0.8]{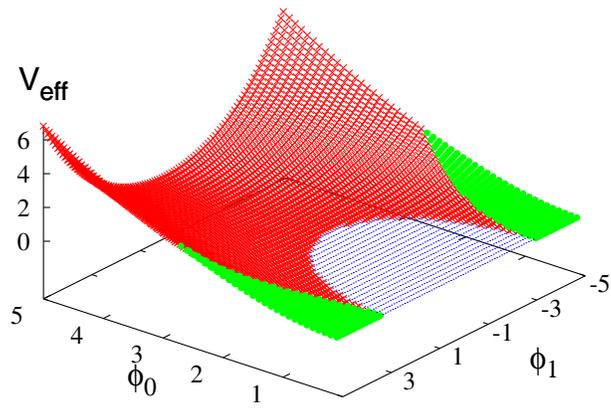}
\vspace*{6mm}
\caption{\label{figure:effpot} The equilibrium effective
potential for $\kappa=0.1$.}
\end{center}
\end{figure} 

\end{document}